\documentclass[fleqn,usenatbib]{mnras}

\usepackage{newtxtext,newtxmath}

\usepackage[T1]{fontenc}
\usepackage{ae,aecompl}

\usepackage{graphicx}	
\usepackage{amsmath}	
\usepackage{amssymb}	

\title[Atmosphere loss in planet-planet collisions]{Atmosphere loss in planet-planet collisions}

\author[T. R. Denman et al.]{Thomas R. Denman$^{1}$\thanks{E-mail: tom.denman@bristol.ac.uk},
Zoe M. Leinhardt$^{1}$,
Philip J. Carter$^{2}$ and
\newauthor Christoph Mordasini$^{3}$
\smallskip
\\
$^{1}$School Of Physics, H.H. Wills Physics Laboratory, University of Bristol, Bristol BS8 1TL, UK\\
$^{2}$Department of Earth and Planetary Sciences, University of California, Davis, One Shields Avenue, Davis, CA, 95616, USA\\
$^{3}$Physics Institute, University of Bern, Gesellschaftsstrasse 6, CH-3012 Bern, Switzerland
}
\date{Accepted XXX. Received YYY; in original form ZZZ}

\pubyear{2019}

\begin{document}
\label{firstpage}
\pagerange{\pageref{firstpage}--\pageref{lastpage}}
\maketitle

\begin{abstract}
Many of the planets discovered by the Kepler satellite are close orbiting Super-Earths or Mini-Neptunes. Such objects exhibit a wide spread of densities for similar masses. One possible explanation for this density spread is giant collisions stripping planets of their atmospheres. 
In this paper we present the results from a series of smoothed particle hydrodynamics (SPH) simulations of head-on collisions of planets with significant atmospheres and bare projectiles without atmospheres. Collisions between planets can have sufficient energy to remove substantial fractions of the mass from the target planet. 
We find the fraction of mass lost splits into two regimes -- at low impact energies only the outer layers are ejected corresponding to atmosphere dominated loss, at higher energies material deeper in the potential is excavated resulting in significant core and mantle loss. Mass removal is less efficient in the atmosphere loss dominated regime compared to the core and mantle loss regime, due to the higher compressibility of atmosphere relative to core and mantle. 
We find roughly twenty per cent atmosphere remains at the transition between the two regimes. We find that the specific energy of this transition scales linearly with the ratio of projectile to target mass for all projectile-target mass ratios measured. The fraction of atmosphere lost is well approximated by a quadratic in terms of the ratio of specific energy and transition energy. We provide algorithms for the incorporation of our scaling law into future numerical studies.

\end{abstract}




\section{Introduction}

There have been a wealth of exoplanet discoveries over recent years, bringing the total number of confirmed planets to more than 4000.  
About twenty percent have a mass between that of Earth and Neptune (henceforth `Super-Earth mass'). Of these, two thirds orbit at a distance less than that of Mercury from their host stars, and more than three quarters are in systems with at least one other planet \citep{pl_conf-Oct_19} .

Planets in this mass and orbital distance range are very diverse, with measured densities ranging from between 0.03 and 12.7 $\mathrm{g\,cm}^{-3}$. Even planets within the same solar system at similar orbital radii can be vastly dissimilar. The two innermost planets of the Kepler-107 system \citep{Bonomo2019}, for example, both have similar radii ($1.5-1.6\mathrm{R_{\oplus}}$), but Kepler-107b has a density of $5.3\,\mathrm{g\,cm^{-3}}$ compared to Kepler-107c's $12.6\,\mathrm{g\,cm^{-3}}$. This high density implies that Kepler-107c must have a different composition from 107b.

There are two main theories that explain the diversity of densities observed in Super-Earths and Mini-Neptunes:
1) XUV radiation from the central star \citep{Lopez2012,Lopez2013,Owen2013,Jin2014} strips close orbiting planets of some of their lighter elements, leaving them denser.
2) Giant impacts eject the lighter outer material (such as crust or atmosphere) from both bodies leaving a denser remnant planet \citep{Inamdar2016,Bonomo2019}.

XUV radiation cannot always explain large differences in density for planets orbiting at similar semi-major axes in the same planetary system, for example, Kepler-107b and c. 
These two planets orbit at a similar distance from their parent star and have similar physical radii but differ significantly in planetary mass. The outer Kepler-107c is more than twice as dense as the inner-most 107b. Kepler-107c's density can not be explained by XUV radiation because it is orbiting outside of the less massive and also less dense Kepler-107b which would have lost more material due to irradiation. Thus, a giant impact is the best explanation \citep{Bonomo2019}.

Giant impacts have been suggested as the explanation for density diversity in several planetary systems \citep{Liu2015,Inamdar2016}. 
In addition, many of the tightly orbiting high-multiplicity systems detected by Kepler appear to be on the borders of stability \citep{Fang2013}. Therefore, it is not unreasonable to consider the planets gravitationally interacting with one another in such a way that either they eject one or both planets or that they collide with one another. \citet{Barnes2004} showed that small perturbations in orbit will lead to such ejections or collisions. 
\cite{Volk2015} even suggested close orbiting groups of planets are common in the formation of inner solar systems ($a$<0.5 AU), including our own, but that they are not stable long term, and will undergo collisional disruption and consolidation. This suggests that the high multiplicity systems that we observe are either these initial unstable systems, or that the planets are in an arrangement that is stable for long time periods.

Several previous works \citep[e.g.][]{Schlichting2015,Inamdar2016} have demonstrated that giant impacts have the potential to remove large fractions of these planets' atmospheres, leading to substantial density enhancement. 
In this paper we directly calculate atmosphere stripping via 3D modelling of head-on giant impacts between Mini-Neptune mass planets with significant atmospheres and lower mass bare Super-Earth impactors. 

\subsection{Previous Work}

Until recently, simulations of atmospheric losses due to such giant collisions were too computationally expensive to run within a reasonable time frame. Recent advances, however, have meant high resolutions are now much more attainable. 

Because of computational demands much early work focused on analytical predictions (some of which focus on significantly lower mass atmospheres than we consider here, such as Earth-like atmospheres), for example \citet{Genda2003} which discusses how much of an Earth-like protoplanet's atmosphere is likely to survive the giant impact phase. They used one dimensional models to calculate the amount of material jettisoned after a collision from the ground velocity beneath it, showing that for the canonical Moon-forming impact only $\approx20$ per cent of mass would be lost. \citet{Genda2003} also showed that the ground velocity needs to reach the escape velocity for total atmosphere erosion. 
\citet{Schlichting2015} expanded upon this by formulating a method of predicting atmospheric loss from a wide range of projectiles colliding with terrestrial planets. This method is the one used by \citet{Inamdar2016} when discussing the density diversity of Super-Earth mass exoplanets. \citet{Inamdar2016} showed that a single collision between similarly sized exoplanets can cause a decrease in the mass ratio of atmospheric envelope to central core material by a factor of 2, which in turn leads to a density increase of a factor of 2--3. \citet{Inamdar2016} thus suggested giant impacts as a cause for the observed density diversity.

Despite computational limitations some progress has been made in simulating collisions of planets with gaseous envelopes. \citet{Liu2015} considered a model of the Kepler-36 system (with target planets of 6--8\% atmosphere by mass). They found that a collision could cause the density difference observed between the two planets in this system, and suggested that giant collisions might therefore be the cause of the dispersion we observe in the mass-radius relationship for Super-Earth mass planets. \citet{Hwang2017a} and \citet{Hwang2017b} used an $N$-body code to model the evolution and stability of high multiplicity planetary systems (specifically Kepler-11 and Kepler-36). For the collisions they used a smoothed particle hydrodynamics (SPH) code and assumed atmosphere mass fractions of 5--15\% to determine the results of highly grazing collisions (collisions where only the atmospheres overlap), they showed that typically the higher mass core will accrete more of the disrupted gas envelope leading to increasing density contrasts between the two planets. Due to problems with their equation of state however, they where unable to simulate the results of head-on collisions. \citet{Kegerreis2018} also used SPH to model collisions involving targets with atmospheres, to see if they could model the formation of Uranus' off axis rotation and unusual magnetic field. Unlike this paper, \citet{Kegerreis2018} focus mostly on higher impact parameter collisions in order to study the change in rotation. They considered ice giants as opposed to the metal and silicate planets with atmospheres studied in this work.

In this paper we present the results from a series of head-on collisions between Super-Earth mass planets where each of the three different mass targets is a Mini-Neptune with a significant hydrogen envelope (8--33\%). We run a large series of numerical simulations with a wide array of atmosphere-less Super-Earth projectiles. We provide scaling relations for material loss applicable to a wide range of masses that can be used in $N$-body simulations and population synthesis models.

\section{Methods}

\subsection{Numerical code}

The simulations presented in this paper were run using the SPH code GADGET-2 \citep{Springel2005}. Although GADGET-2 was initially designed for simulations on cosmological scales, we have used a modified version to model our planets that includes tabulated equations of state for the planetary constituents. For further detail on these modifications see \citet{Marcus2009} and \citet{Matija2012}. We further modified GADGET-2 to include an ideal gas atmosphere component.

In SPH codes, such as GADGET-2, the material is split into separate particles each representing an ensemble of material. The continuous fluid properties, such as the density, for each ensemble of particles are calculated using kernel interpolation methods. The gravitational forces on the other hand are calculated using hierarchical tree methods \citep{Springel2005}.
We ran GADGET-2 in `Newtonian' mode with timestep synchronisation, and the standard relative cell-opening criterion. We use the standard timestep criterion as described in \citet{Springel2005}, where the smaller of the timestep based on the gravitational softening and the acceleration, or the courant condition is used (with a courant factor of 0.1). We use the standard artificial viscosity formulation for Gadget-2 as described in \citet{Springel2005}, with a strength parameter of 0.8.

We modelled the planets as two or three material systems, each planet (both projectile and target) consisting of iron core, forsterite (silicate) mantle, and a hydrogen atmosphere for the target only. In a similar fashion to prior studies \citep{Matija2012,Marcus2009}, the mantle was twice the mass of core. We modelled the atmosphere as a monatomic ideal gas for simplicity and ease of comparison (see section \ref{sec:caveats}). Tabulated ANEOS/MANEOS equations of state (EOS) from \citet{Melosh2007} were used for the iron and forsterite (these tables are available from \citealt{GADGET2EOS2019}).

For the initial hydrogen atmosphere mass, we used the results of the Bern global planet formation model \citep{Alibert2005, Mordasini2018} which was based on the core accretion paradigm. The model calculated the accretion of H/He of forming planets embedded in a protoplanetary disk by solving the 1D spherically symmetric interior structure equations \citep{Pollack1996}. In these calculations, the grain opacity in the protoplanetary atmospheres was reduced by a factor 0.003 relative to interstellar medium grain opacities following \citet{Mordasini2014a} so it has a value consistent with observed metal enrichment of giant planets. Other effects considered were the accretion of planetesimals, orbital migration, disk evolution, and $N$-body interactions. These models predicted that for planets with masses between 1 and 7 $\mathrm{M_{\oplus}}$, the mass of the H/He envelope, $M_\mathrm{atmos}$, at the end of the disk lifetime can be approximated by
\begin{equation}
    \frac{M_{\mathrm{atmos}}}{\mathrm{M_{\oplus}}}=0.01\times\left(\frac{M_{\mathrm{c\&m}}}{\mathrm{M_{\oplus}}}\right)^{3}
    \label{eq:atmos_frac_pred}
\end{equation}
where $M_{\mathrm{c\&m}}$ is the combined mass of iron core and forsterite mantle. This yields an envelope mass of $M_{\mathrm{atmos}}=1.25$ $\mathrm{M_{\oplus}}$ for a combined mantle and core mass of of $M_{\mathrm{c\&m}}=5$ $\mathrm{M_{\oplus}}$ and $M_{\mathrm{atmos}}=10$ $\mathrm{M_{\oplus}}$ for 10 $\mathrm{M_{\oplus}}$ of core and mantle. A 10 $\mathrm{M_{\oplus}}$ core is close to the critical mass for runaway gas accretion which sets in when $M_{\mathrm{c\&m}} \approx M_{\mathrm{atmos}}$, which is approximately captured by this relation. One should, however, note that in the Bern model simulations, a large spread around this mean relation of about one order in magnitude is observed.

\subsection{Hardware}

The simulations were each run using a full node on the University of Bristol's Bluecrystal supercomputers, either on phase 3 or phase 4. Phase 3 nodes consist of 16 core 2.6 GHz SandyBridge processors with 59.7 GiB RAM altogether  \citep{BlueCrystal3}, whilst phase 4 nodes have two 14 core 2.4 GHz Intel E5-2680 v4 (Broadwell) CPUs with the whole node having a combined 128 GiB of RAM \citep{BlueCrystal4}. Each collision took between half a day and a day depending on the masses and impact energies involved.

\subsection{Initial conditions}

To generate the initial planets, we began by creating the central core and mantle. To generate density profiles we used radial temperature profiles \citep[from][]{Valencia2006} as well as estimates for the average bulk density of each type of material (i.e core or mantle) and the radial range that the core and mantle occupies. From this initial assumption of constant density per material layer we iteratively generated new density profiles using gravitational and hydrostatic pressure calculations, until we obtained a density profile that was consistent with our equation of state, the expected hydrostatic pressure and the input temperature profile.    

Once a consistent profile was obtained it was then used to generate the position of each of the particles by splitting the planet into a number of radial bins based on its mass and randomly positioning a number of particles within each bin proportional to the bin's mass. The final number of bins being adjusted so that we could reach our desired mass to a tolerance of 1 per cent. The type of material being added to each bin was decided by the material ranges given for the initial density profile.
After generating an initial SPH planet it was then equilibrated in isolation for $10^5\mathrm{s}$ (simulated seconds) using a preliminary run of GADGET-2, to ensure that we were running simulations with a planet that was stable. During equilibration we used two `artificial cooling' methods, velocity damping of the particles, i.e applying a restitution factor of 50\% each timestep \citep[see][]{Carter2018}, and also entropy forcing, the entropy of each particle is reset to a constant value for each material at every timestep ($1.3\,\mathrm{kJ\,K^{-1}\,kg^{-1}}$ for the iron core, $3.2\,\mathrm{kJ\,K^{-1}\,kg^{-1}}$ for the mantle). This entropy forcing ensures that we produce planets with isentropic layers.

To generate a planet with an atmosphere, we added an atmosphere to the previously generated core and mantle only planets. 
The radial profiles of the planetary atmospheres were generated with the planet interior structure and evolution model \texttt{completo21}, which has already been described in several publications \citep{mordasini2012,Jin2014,Linder2019}. Therefore we only give a short overview here. 
The structure of the atmosphere was found in the 1D spherically symmetric approximation by solving the usual equations of mass conservation, hydrostatic equilibrium, energy generation, and energy transport 
\begin{alignat}{2}
\frac{\partial m}{\partial r}&=4 \pi r^{2} \rho    &\quad  \quad \frac{\partial P}{\partial r}&=-\frac{G m}{r^{2}}\rho    \\
\frac{\partial l}{\partial r}&=0            & \frac{ \partial T}{\partial r}&=\frac{T}{P}\frac{\partial P}{\partial r}\nabla(T,P)          
\end{alignat}
where $m$ is the mass inside of a radius (distance to the planet's centre) $r$, $\rho$ the gas density, $P$ the pressure, $G$ the gravitational constant, $l$ the (intrinsic) luminosity, $T$ the temperature, and $\nabla(T,P)$ the temperature gradient.
The Schwarzschild criterion was used to decide whether the energy transport occurs in a layer via radiative diffusion or convection, meaning that $\nabla$ is always the smaller of the radiative and the adiabatic gradient.  When solving the structure equations, we assumed opacities corresponding to a condensate-free gas of solar composition \citep{Freedman2014}, and, in contrast to past publications, an ideal gas EOS. 
For the (intrinsic) luminosity, $l$,  of the planets, which needs to be specified in order to solve the structure equations, we employed a simple power law scaling with planet mass. Like equation \ref{eq:atmos_frac_pred}, the luminosity scaling was also based on formation simulations with the Bern model. The results obtained correspond to planet ages of 10 Myr. 
This age corresponds to a time when the systems are still dynamically active (soon after the dispersion of the eccentricity damping gas disk), so many collisions should occur. This gives, 
\begin{equation}
L/\mathrm{L_{J}} \simeq 0.1\times\left(\frac{M_{\mathrm{planet}}}{\mathrm{M_{\oplus}}}\right)^{1.5},
\end{equation}
where $\mathrm{L_{J}}$ is Jupiter's luminosity and $M_{\mathrm{planet}}$ is the total mass of the planet (again there is scatter around this relation).

We then used this radial mass profile for the atmosphere to determine the position of atmosphere particles, by splitting the profile into radial bins and placing particles proportionally to the mass of the bins at random positions within them, in a similar fashion to the core and mantle. This new body was equilibrated for a longer time of 4--8$\times10^5\,\mathrm{s}$ until the radius of the planet had converged to a constant value. The pseudo-entropy of the atmosphere was forced to a value of $5\times10^{11}\,\mathrm{Ba\,g^{-\gamma}\,cm^{3\gamma}}$ where the adiabatic index was $\gamma=5/3$, this ensured that the atmosphere would not reach densities where it would sink underneath the mantle and core material, but also that the base of the ideal gas atmosphere was as close as possible to our predicted temperatures.

For the core and mantle of our targets we used a resolution of $10^{5}$ particles. All other particles in each simulation were made the same mass as the core and mantle particles, resulting in a total resolution between $1.2\times10^{5}$ and $2.5\times10^{5}$ particles.  This results in an atmosphere `thickness' of between 5 and 10 layers of particles. Our total particle number is smaller than that suggested by \citet{Kegerreis2019}, however, we are interested in large changes in mass of the largest remnant so our resolution should be sufficient. 
We have run resolution tests of head-on collisions of 5 $\mathrm{M_{\oplus}}$ planets with 0.9 $\mathrm{M_{\oplus}}$ atmospheres against one another at both half and double our standard resolution (of $10^{5}$ particles in the target core and mantle) at impact velocities of 20 and 40 $\mathrm{km\,s^{-1}}$. The key quantities we measure: mass of the largest remnant, atmosphere loss fraction and core and mantle loss fraction, each vary from the mean value at that velocity by less than 5\% except for the core and mantle loss fraction at low velocity where only a very small number of particles are being lost. We did not consider losses or remnants of only a few hundred particles resolvable with SPH methods.

\subsubsection{The Point of Impact}

In collisional studies impact parameters such as velocity, impact energy etc. are normally  measured in terms of the point of first physical contact between the two planetary bodies \citep[e.g.][]{Leinhardt2012}. For planets with atmospheres this becomes more complex, however, as atmosphere densities decay approximately exponentially so there is no clear boundary at which the atmosphere ends. Tidal forces between planets also have a stronger distorting effect on the atmosphere than the core and mantle.  Thus we use the time when mantle surfaces touched as our point of impact because this can be clearly defined.

To obtain an initial start position from our desired collision parameters (for example velocity and impact angle), the following process was used: 1) the two planets were represented by point masses at their centres, these two planets were placed at their position at the point of impact and set at the predicted velocity; 2) Time reversal symmetry was then used to trace the path of the projectile back to a separation of five times the sum of the projectile and target mantle radii (excluding atmosphere). To determine the projectiles path we used a simple Verlet integrator, set the target planet to be centred on the origin, and calculated the acceleration as given by the relative gravitational force between two point particles. The choice of starting separation was a somewhat arbitrary one intended to reduce the tidal forces compared to the starting distances of previous similar studies due to us equilibrating planets in isolation and atmospheres being more easy to tidally distort.

\begin{figure}
	\includegraphics[width=\columnwidth]{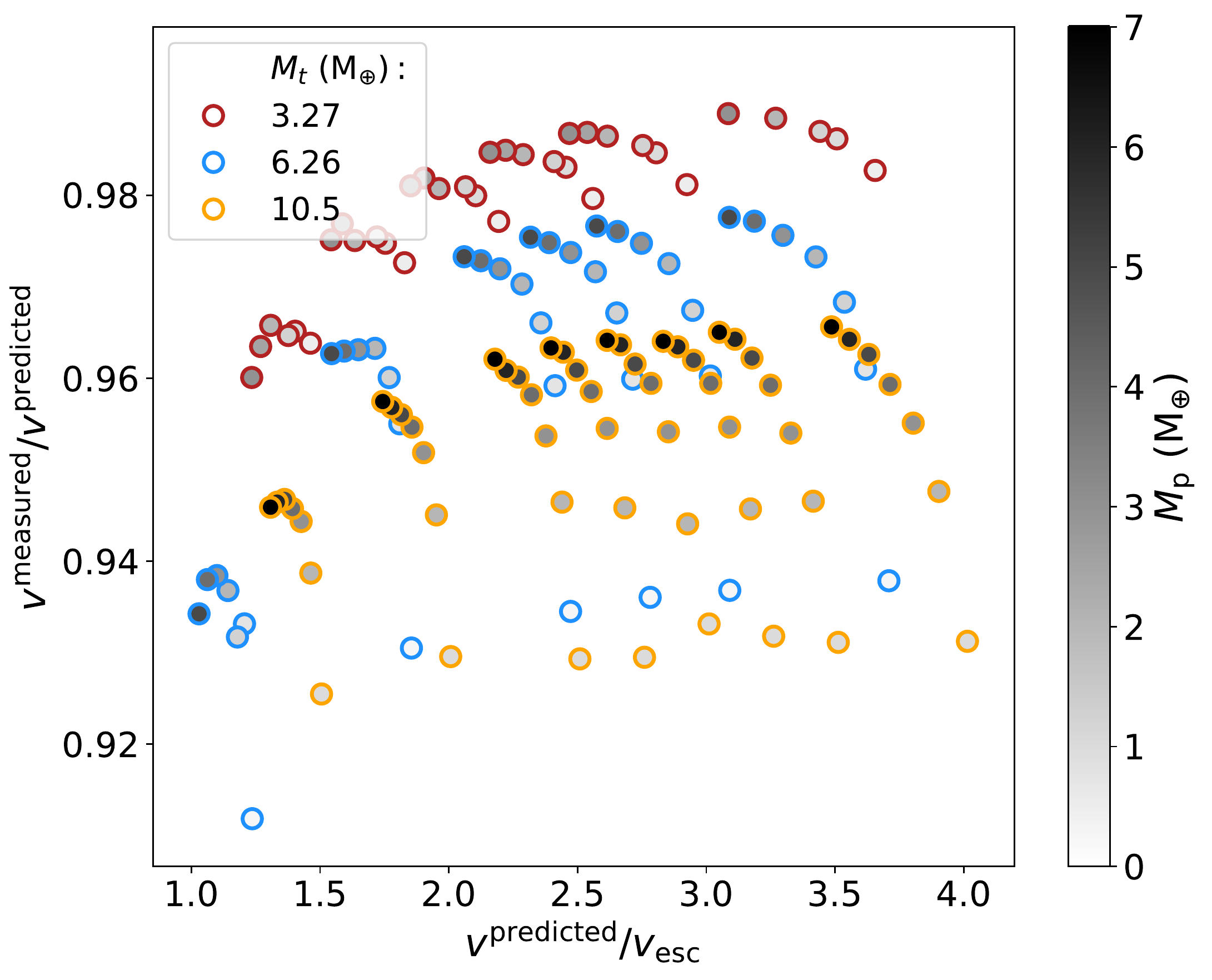}
    \caption{Comparisons of the predicted velocity for the three targets and the ratio between measured and predicted impact velocities. Impact velocities are given in terms of the mutual escape velocity ($v_{\mathrm{esc}}=\left(2GM_{\mathrm{tot}}/R^{'}\right)^{\frac{1}{2}}$ where $R^{'}=\left(3M_{\mathrm{tot}}/4\pi\rho_{\mathrm{bulk}}\right)^{\frac{1}{3}}$ and $\rho_{\mathrm{bulk}}$ is the bulk density of the simulated target). Edge colours show target mass while central colours show projectile mass. The reduced velocity compared to the prediction can be considered a measure of the drag caused by the atmosphere. As might be expected, the denser, higher mass atmospheres around the larger targets tend to cause more drag, the lower mass and lower velocity projectiles tend to experience proportionally more drag as well due to their lower momenta. }
    \label{fig:vact_V_vpred}
\end{figure}

Note using the point where mantles touch as our point of impact is not without its drawbacks. As can be observed in Figure \ref{fig:vact_V_vpred} the presence of the atmosphere causes a significant observable slowing of the relative impact velocity. This slowing is due to atmospheric drag which distorts the projectile planet. For some collisions this distortion is even present when the leading edge of the projectile is greater than an atmospheric scale height from the target. In further calculations, which use impact energy, we used the measured velocity instead of the predicted input value. To measure this collision point precisely we re-simulated the point of impact for our collisions with a higher output frequency (snapshots were taken every 3 seconds as opposed to every 100).

\subsubsection{Input Parameters}

All collisions presented in this paper were head-on, involving a projectile with no atmosphere and a target with atmosphere. Head-on collisions were chosen as they are the most energetic and are, therefore, expected to be the most efficient at removing material \citep{Leinhardt2012}. The three target masses simulated were: a) a $3.0\,\mathrm{M_{\oplus}}$ core and mantle with a $0.27\,\mathrm{M_{\oplus}}$ atmosphere (Table \ref{tab:3_0.27}), b) a $5.01\,\mathrm{M_{\oplus}}$ core and mantle with a $1.25\,\mathrm{M_{\oplus}}$ atmosphere (Table \ref{tab:5_1.25}), and c) a $7.07\,\mathrm{M_{\oplus}}$ core and mantle with a $3.43\,\mathrm{M_{\oplus}}$ atmosphere (Table \ref{tab:7_3.43}). Projectiles had masses between $0.05$ to $1$ times the core and mantle mass of the target, giving 6--7 mass ratios per target distributed approximately evenly between 0.04 and 0.92. These masses were chosen to sample a wide range of parameter space with a limited number of runs. 7--8 collision velocities per target were distributed approximately evenly between $20$ and $80\,\mathrm{km\,s^{-1}}$. These velocities were 1--4 times the mutual escape speed, which we define in a similar fashion to \citet{Leinhardt2012}, 
\begin{equation}
v_{\mathrm{esc}}=\left(2GM_{\mathrm{tot}}/R^{'}\right)^{\frac{1}{2}}, 
\end{equation}
where $M_{\mathrm{tot}}$ is total mass and $R^{'}$ is the radius of an spherical body of mass $M_{\mathrm{tot}}$ and the same density as the bulk density of the simulated target. This measure for mutual escape velocity was chosen due to it being a minimum velocity at which we might expect to see ejection of material. This velocity range meant that we could sample well both collision regimes detailed in section \ref{sec:col_regimes}, as well as the transition between them.

\subsection{Run Parameters}

To determine the length of time we needed for our simulations (the run time) we used the dynamical time for gravitational interactions $t_{\mathrm{dyn}}$. A dynamical time for a process is a prediction of the rough timescale it will take to finish, generated from a few system parameters. For gravity dominated systems the dynamical time is given by: 
\begin{equation}
    t_{\mathrm{dyn}}=\sqrt{\frac{R^3}{GM}}
\end{equation}
Taking the largest initial separation as $R$ and the smallest total mass as $M$, the largest dynamical time we obtain is $~27000$\,s. We elected to use a run time of $10^5$\,s, a value slightly under 4 dynamical times. Whilst a longer run time gives a more reliable estimate of the long term final state after a collision, we were constrained by computation time, and the results we were most interested in (bound mass for each material component of our planets) had already converged to a constant value by this time for all our simulations.

\subsection{Analysis Methods}

\subsubsection{Determining Bound Material}

To determine what material was bound in the largest post-collision remnant we used the same iterative method as \citet{Marcus2009} and \citet{Carter2018}. We began by locating the particle closest to the potential minimum and using kinetic energy and gravitational potential to determine which other particles were gravitationally bound to this deepest particle. Using the total mass and centre of mass of these particles as our new seed, we then iteratively ran through the process of determining which extra particles were bound to the seed, and then adding them to the seed, until either the change in mass of the seed was below a set tolerance or a maximum number of iterations had been reached. For our simulations we obtained single particle differences in mass within a few iterations so our tolerance was set to a single particles mass, and the maximum number of iterations was never reached. Material type was also tracked, to determine the mass of each type of material which was unbound or part of the largest remnant.

\begin{table*}
\caption{
Simulation parameters and results for head-on collisions between a $6.26\,\mathrm{M}_{\oplus}$ target, with a mantle surface radius of $1.49\,\mathrm{R}_{\oplus}$, and an atmosphere scale height of $0.60\,\mathrm{R}_{\oplus}$, and a projectile of mass $M_{\mathrm{p}}$ and radius $R_{\mathrm{p}}$. The first digit of the collision ID denotes the non-atmospheric target mass ($M_\mathrm{t}^{\mathrm{core}} = 5.01 \, \mathrm{M}_{\oplus}$, therefore, $M_\mathrm{t}^{\mathrm{atmos}} = 1.25 \, \mathrm{M}_{\oplus}$). $v_{\mathrm{init}}$ is the initial relative speed of the projectile with respect to the target, at an initial separation of $S$ to give a predicted impact speed of $v_{\mathrm{imp}}^{\mathrm{pred}}$, the measured velocity at the point where the mantles touch is $v_{\mathrm{imp}}^{\mathrm{meas}}$. $M_{\mathrm{LR}}$ is the total mass of the largest post-collision remnant, where $M_{\mathrm{LR}}^{\mathrm{atmos}}$ and $M_{\mathrm{LR}}^{\mathrm{core}}$ are the atmospheric and non-atmospheric masses, respectively, the final column gives the category we have given that collision. We have used '-' to denote final snapshots where there where too few particles in the largest remnant to be able to properly resolve. Simulation data for $M_\mathrm{t} = 3.27 \, \mathrm{M}_{\oplus}$ and $M_\mathrm{t} = 10.5 \, \mathrm{M}_{\oplus}$ can be found in Tables \ref{tab:3_0.27} \& \ref{tab:7_3.43}. }
\label{tab:5_1.25}

\begin{tabular}{lcccccccccccccccc}
\hline
ID&$M_{\mathrm{p}}$ &$R_{p}$ &$v_{\mathrm{imp}}^{\mathrm{pred}}$ &$\frac{v_{\mathrm{imp}}^{\mathrm{pred}}}{v_{\mathrm{esc}}}$ &$v_{\mathrm{imp}}^{\mathrm{meas}}$ &$\frac{v_{\mathrm{imp}}^{\mathrm{meas}}}{v_{\mathrm{esc}}}$ &$v_{\mathrm{init}}$ &$\frac{v_{\mathrm{init}}}{v_{\mathrm{esc}}}$ &$S$ &$M_{\mathrm{LR}}$ &$M_{\mathrm{LR}}^{\mathrm{atmos}}$ &$M_{\mathrm{LR}}^{\mathrm{core}}$ & $\frac{M_{LR}}{M_{\mathrm{tot}}}$ & $X_{\mathrm{loss}}^{\mathrm{atmos}}$ & $X_{\mathrm{loss}}^{\mathrm{c\&m}}$ & Category \\
&$\mathrm{M}_{\oplus}$ &$\mathrm{R}_{\oplus}$ &$\mathrm{km}\,\mathrm{s}^{-1}$ & &$\mathrm{km}\,\mathrm{s}^{-1}$ & &$\mathrm{km}\,\mathrm{s}^{-1}$ & &$\mathrm{R}_{\oplus}$ &$\mathrm{M}_{\oplus}$ &$\mathrm{M}_{\oplus}$ &$\mathrm{M}_{\oplus}$& & & & \\
\hline
5-0& 0.25& 0.62& 20.00& 1.24& 18.24& 1.13& 9.38& 0.58& 11.15& 6.50& 1.24& 5.26& 1.0& 0.01& -0.0& AM-CM\\
5-1& 0.25& 0.62& 30.00& 1.85& 27.92& 1.73& 24.25& 1.50& 11.16& 6.43& 1.17& 5.26& 0.99& 0.06& -0.0& AL-CM\\
5-2& 0.25& 0.62& 40.00& 2.47& 37.38& 2.31& 35.89& 2.22& 11.15& 6.30& 1.05& 5.26& 0.97& 0.16& -0.0& AL-CM\\
5-3& 0.25& 0.62& 45.00& 2.78& 42.12& 2.60& 41.39& 2.56& 11.18& 6.22& 0.97& 5.25& 0.96& 0.22& 0.0& AL-CM\\
5-4& 0.25& 0.62& 50.00& 3.09& 46.84& 2.90& 46.77& 2.89& 11.15& 6.14& 0.89& 5.25& 0.94& 0.29& 0.0& AL-CM\\
5-5& 0.25& 0.62& 60.00& 3.71& 56.27& 3.48& 57.34& 3.54& 11.18& 5.94& 0.71& 5.23& 0.91& 0.43& 0.01& AL-CM\\
5-6& 0.75& 0.88& 20.00& 1.21& 18.66& 1.13& 10.22& 0.62& 11.93& 6.90& 1.13& 5.76& 0.98& 0.1& 0.0& AL-CM\\
5-7& 0.75& 0.88& 30.00& 1.81& 28.65& 1.73& 24.58& 1.48& 11.93& 6.66& 0.91& 5.75& 0.95& 0.27& 0.0& AL-CM\\
5-8& 0.75& 0.88& 40.00& 2.41& 38.37& 2.31& 36.11& 2.18& 11.96& 6.29& 0.56& 5.73& 0.9& 0.55& 0.01& AL-CM\\
5-9& 0.75& 0.88& 45.00& 2.71& 43.20& 2.61& 41.58& 2.51& 11.94& 6.11& 0.41& 5.69& 0.87& 0.67& 0.01& AL-CM\\
5-10& 0.75& 0.88& 60.00& 3.62& 57.66& 3.48& 57.48& 3.47& 11.96& 4.62& 0.07& 4.55& 0.66& 0.94& 0.21& AL-CE\\
5-11& 1.25& 1.02& 20.00& 1.18& 18.63& 1.10& 10.11& 0.60& 12.35& 7.28& 1.02& 6.26& 0.97& 0.18& 0.0& AL-CM\\
5-12& 1.25& 1.02& 30.00& 1.77& 28.80& 1.70& 24.54& 1.45& 12.35& 6.92& 0.68& 6.25& 0.92& 0.46& 0.0& AL-CM\\
5-13& 1.25& 1.02& 45.00& 2.65& 43.52& 2.57& 41.56& 2.45& 12.37& 5.86& 0.19& 5.68& 0.78& 0.85& 0.09& AL-CA\\
5-14& 1.25& 1.02& 50.00& 2.95& 48.37& 2.85& 46.93& 2.77& 12.36& 5.04& 0.09& 4.95& 0.67& 0.93& 0.21& AL-CE\\
5-15& 1.25& 1.02& 60.00& 3.54& 58.10& 3.42& 57.46& 3.39& 12.36& 2.94& 0.00& 2.94& 0.39& 1.0& 0.53& TAL-CE\\
5-16& 2.00& 1.18& 20.00& 1.14& 18.74& 1.07& 9.66& 0.55& 12.81& 7.89& 0.91& 6.99& 0.95& 0.27& 0.0& AL-CM\\
5-17& 2.00& 1.18& 30.00& 1.71& 28.90& 1.65& 24.36& 1.39& 12.82& 7.49& 0.53& 6.96& 0.91& 0.58& 0.01& AL-CM\\
5-18& 2.00& 1.18& 40.00& 2.28& 38.81& 2.22& 35.96& 2.05& 12.83& 6.42& 0.21& 6.22& 0.78& 0.83& 0.11& AL-CA\\
5-19& 2.00& 1.18& 45.00& 2.57& 43.72& 2.50& 41.45& 2.37& 12.81& 5.48& 0.09& 5.38& 0.66& 0.93& 0.23& AL-CA\\
5-20& 2.00& 1.18& 50.00& 2.85& 48.63& 2.78& 46.83& 2.67& 12.84& 4.15& 0.01& 4.14& 0.5& 0.99& 0.41& TAL-CE\\
5-21& 2.00& 1.18& 60.00& 3.43& 58.40& 3.33& 57.39& 3.28& 12.83& 1.12& 0.00& 1.12& 0.14& 1.0& 0.84& TAL-CE\\
5-22& 3.01& 1.32& 20.00& 1.10& 18.77& 1.03& 8.69& 0.48& 13.24& 8.73& 0.81& 7.92& 0.94& 0.35& 0.01& AL-CM\\
5-23& 3.01& 1.32& 30.00& 1.65& 28.89& 1.59& 23.99& 1.32& 13.25& 8.38& 0.49& 7.89& 0.9& 0.61& 0.02& AL-CM\\
5-24& 3.01& 1.32& 40.00& 2.20& 38.88& 2.14& 35.71& 1.96& 13.25& 6.76& 0.17& 6.59& 0.73& 0.86& 0.18& AL-CA\\
5-25& 3.01& 1.32& 45.00& 2.47& 43.82& 2.41& 41.24& 2.27& 13.25& 5.45& 0.07& 5.39& 0.59& 0.94& 0.33& AL-CA\\
5-26& 3.01& 1.32& 50.00& 2.75& 48.74& 2.68& 46.64& 2.56& 13.26& 3.78& 0.00& 3.78& 0.41& 1.0& 0.53& TAL-CE\\
5-27& 3.01& 1.32& 60.00& 3.30& 58.54& 3.22& 57.23& 3.15& 13.28& -& -& -& -& 1.0& -& SCD\\
5-28& 4.01& 1.43& 20.00& 1.06& 18.76& 1.00& 7.47& 0.40& 13.58& 9.66& 0.81& 8.85& 0.94& 0.35& 0.02& AL-CM\\
5-29& 4.01& 1.43& 30.00& 1.59& 28.89& 1.53& 23.57& 1.25& 13.60& 9.37& 0.51& 8.86& 0.91& 0.59& 0.02& AL-CM\\
5-30& 4.01& 1.43& 40.00& 2.12& 38.91& 2.07& 35.43& 1.88& 13.60& 7.46& 0.17& 7.29& 0.73& 0.86& 0.19& AL-CA\\
5-31& 4.01& 1.43& 45.00& 2.39& 43.87& 2.33& 41.00& 2.18& 13.59& 5.95& 0.06& 5.89& 0.58& 0.95& 0.35& TAL-CA\\
5-32& 4.01& 1.43& 50.00& 2.66& 48.80& 2.59& 46.43& 2.47& 13.60& 4.05& 0.00& 4.05& 0.39& 1.0& 0.55& TAL-CE\\
5-33& 4.01& 1.43& 60.00& 3.19& 58.63& 3.11& 57.06& 3.03& 13.59& 0.30& 0.00& 0.30& 0.03& 1.0& 0.97& SCD\\
5-34& 5.01& 1.52& 20.00& 1.03& 18.68& 0.96& 5.80& 0.30& 13.84& 10.72& 0.79& 9.93& 0.95& 0.37& 0.01& AL-CM\\
5-35& 5.01& 1.52& 30.00& 1.54& 28.88& 1.49& 23.10& 1.19& 13.85& 10.38& 0.53& 9.85& 0.92& 0.58& 0.02& AL-CM\\
5-36& 5.01& 1.52& 40.00& 2.06& 38.93& 2.00& 35.12& 1.81& 13.84& 8.34& 0.20& 8.14& 0.74& 0.84& 0.19& AL-CA\\
5-37& 5.01& 1.52& 45.00& 2.32& 43.89& 2.26& 40.73& 2.10& 13.85& 6.93& 0.08& 6.85& 0.62& 0.94& 0.32& AL-CA\\
5-38& 5.01& 1.52& 50.00& 2.57& 48.83& 2.51& 46.19& 2.38& 13.86& 4.80& 0.00& 4.80& 0.43& 1.0& 0.52& TAL-CE\\
5-39& 5.01& 1.52& 60.00& 3.09& 58.65& 3.02& 56.86& 2.93& 13.88& 0.93& 0.00& 0.93& 0.08& 1.0& 0.91& SCD\\
\hline
\end{tabular}

\end{table*}

\subsubsection{Collision categorisation}

To categorise the collision outcomes for our data we consider separately atmosphere (`A') and core and mantle material (`C'). For atmospheres if there was greater than 95 per cent of the initial atmosphere mass in the final remnant we considered there to be a merger (`AM'), on the other hand if there was less than 5 per cent remaining we categorised this as total loss (`TAL'), the rest were considered to undergo partial loss (`AL'). Although hydrogen atmospheres as small as $0.1-1\%$ of a planet's total mass can have a significant effect on its radius, we do not have the resolution to accurately probe atmosphere mass losses that small. For core and mantle, we define mergers (`CM') for $>95$ per cent of the total mass of core and mantle from both projectile and target remaining in the largest remnant. If the mass of core and mantle in the largest remnant is greater than that initially in the target we have an accretion event (`CA'), if it is less we have an erosion event (`CE'). If there was less than $10$ per cent of the total mass remaining in the largest remnant we define it as a super-catastrophic disruption (`SCD'). Tables \ref{tab:5_1.25}, \ref{tab:3_0.27} and \ref{tab:7_3.43}, categorise our final results for each collision using this system.

\begin{figure*}
	\includegraphics[width=\linewidth]{{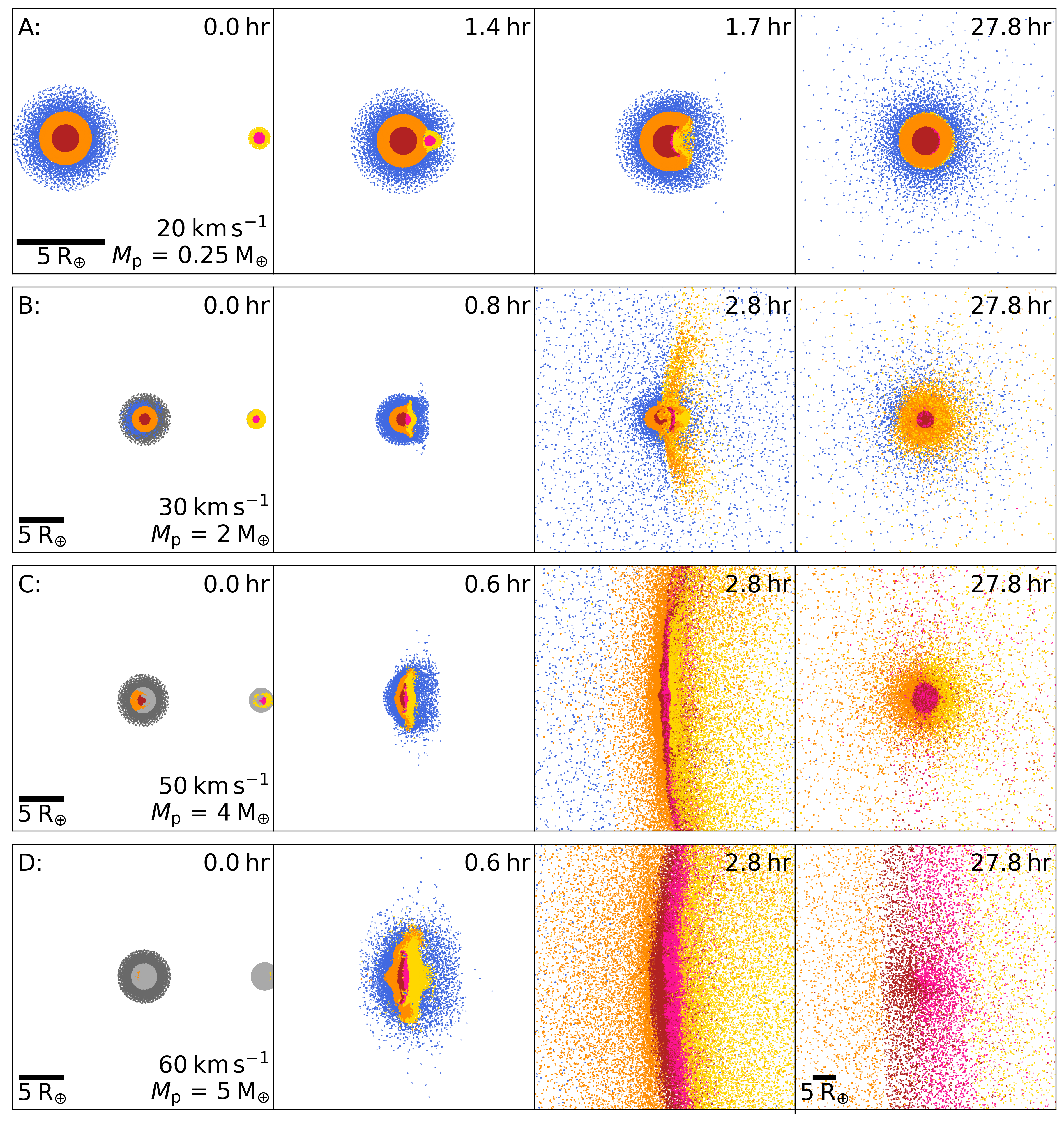}}
    \caption{Cross-sectional snapshots sliced through the midplane for a series of head-on collisions showing different collision outcomes from merging to catastrophic disruption as the specific relative kinetic energy $Q_\mathrm{R}$ increases from row A to row D. Colour denotes material, red and pink -- iron core, orange and yellow -- forsterite mantle, and blue -- hydrogen atmosphere. The additional colours in the first panel denote material that will not be bound by the end of the simulation, black being atmosphere, and grey core or mantle.  All collisions shown have the same target mass ($M_\mathrm{t} = 6.26\,\mathrm{M}_{\oplus}$) but differ in projectile mass ($M_\mathrm{p}$) and impact speed ($\mathbf{v_\mathrm{imp}}$):  A) Atmosphere and core and mantle merger -- $\mathbf{v_\mathrm{i}} = 20$ km s$^{-1}$,  $M_\mathrm{p}=0.25\,\mathrm{M}_{\oplus}$ (Table \ref{tab:5_1.25} 5-0); B) Atmosphere loss and core and mantle merger-- $\mathbf{v_\mathrm{i}} = 30$ km s$^{-1}$, $M_{\mathrm{p}}=2\,\mathrm{M}_{\oplus}$ (Table \ref{tab:5_1.25} 5-17); C) Total atmosphere loss and core and mantle erosion -- $\mathbf{v_\mathrm{i}} = 50$ km s$^{-1}$, $M_\mathrm{p}=4\,\mathrm{M}_{\oplus}$ (Table \ref{tab:5_1.25} 5-32); D) Supercatastrophic disruption $\mathbf{v_\mathrm{i}} = 60$ km s$^{-1}$, $M_\mathrm{p}=5\,\mathrm{M}_{\oplus}$ (Table \ref{tab:5_1.25} 5-39).
    Post collision remnants were inflated in comparison to the initial planets and the expected radius of the resultant planet. This `puffiness' is because we do not cool our final remnants until they reach equilibrium, we only run the simulations until the mass of bound material converges. Videos of these 4 collisions can be found in the online supplementary material.  
    }
    \label{fig:Snapshots}
\end{figure*}

\section{Results}

The main aim of this paper is to determine scaling laws for atmosphere loss as well as total material loss during head-on giant impacts. The wide breadth of impact energies that are simulated in this work have uncovered a broad range of outcomes from near perfect merging events to highly energetic catastrophic disruption (see Figure \ref{fig:Snapshots}).

In the process of determining the loss scaling laws we have found that the atmosphere loss, core and mantle material loss, and the largest remnant mass are well behaved functions of specific impact energy: $Q_\mathrm{R} = \frac{1}{2}\mu V_{\mathrm{imp}}^2 / M_\mathrm{tot}$, where $\mu$ is the reduced mass, $V_{\mathrm{imp}}$ is the impact velocity, and $M_\mathrm{tot}$ is the sum of the projectile and target masses \citep[see][]{Leinhardt2012}. We find that atmosphere dominates the material lost until the impact energy is large enough to remove more than 80\% of the atmosphere at which point mantle and core material begin to be removed significantly as well.  It is also at this point that there is a break and steepening in slope of the largest post collision remnant mass ($M_{\mathrm{lr}}$) as a function of $Q_\mathrm{R}$ (Figure \ref{fig:QrMultipleTarget}). In other words, the atmosphere of the target planet cannot be completely removed in one giant impact without significantly eroding the planet.

\subsection{Mass Loss}
\begin{figure}
	
	\includegraphics[width=\columnwidth]{{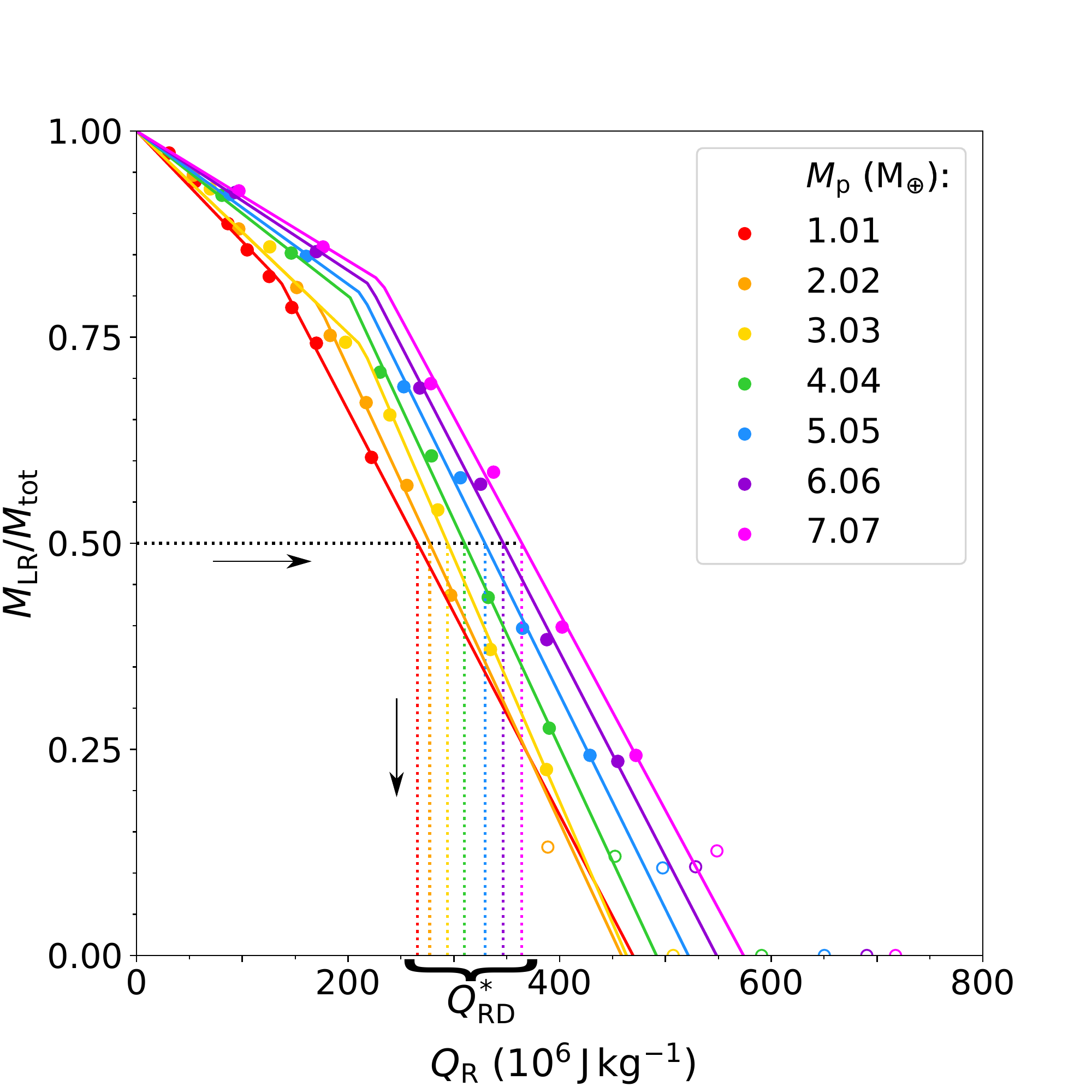}}
    \caption{A comparison of the mass of the largest remnant for each collision compared with its specific impact energy, overlaid with a graphical representation of the process used to determine $Q_{\mathrm{RD}}^*$. The circles represent the fraction of the total mass which remains in the largest remnant after a collision with a target mass of $M_\mathrm{t} = 10.5 \,\mathrm{M}_{\oplus}$ as a function of specific impact energy $Q_\mathrm{R}$. Each colour indicates one of seven $M_\mathrm{p}$ values. A filled circle represents a point used for fits, whilst open circles are points with $\frac{M_{\mathrm{LR}}}{M_{\mathrm{tot}}}<0.2$
    which we considered too close to the super-catastrophic disruption regime which our fit is not designed to cover, for all collisions where $<100$ particles were observed in the largest remnant, we also considered the results to be below the resolution limit of the simulation. The solid lines represent our fit to the data for each projectile-target mass ratio, following equation \ref{eq:QRD_finder_fit}. From this fit the empirically determined value of $Q_{\mathrm{RD}}^*$ is shown on the horizontal axis by the intersection of a coloured dotted line matching the $M_\mathrm{p}$ colour and $M_{\mathrm{LR}}/M_{\mathrm{tot}} = 0.5$ on the $y$-axis. }
    
    \label{fig:QRNormalisation}
\end{figure}

\label{sec:col_regimes}
We begin by comparing our simulation results to the prescription described in \citet{Leinhardt2012} based on collisions of planetesimals and planetary embryos constructed using solely rocky material. A key result from \citet{Leinhardt2012} was that the mass of the largest post-collision remnant, $M_{\mathrm{LR}}$, scales with $Q_{\mathrm{R}}$:
\begin{equation}
    \frac{M_{\mathrm{LR}}}{M_{\mathrm{tot}}}=-0.5\left(\frac{Q_\mathrm{R}}{Q_{\mathrm{RD}}^{*}}-1\right)+0.5,
    \label{eq:Leinhardtetal2012}
\end{equation} where $Q_{\mathrm{RD}}^{*}$ is the catastrophic disruption threshold (the energy required to disperse enough material that the largest post-collision remnant is half the total system mass). 
Normalising $Q_{\mathrm{R}}$ by $Q_{\mathrm{RD}}^{*}$ causes the largest remnant mass fraction for different target and projectile masses to fit on the same line.

Figure \ref{fig:QRNormalisation} shows the largest remnant mass from our simulations as a function of specific impact energy for collisions with a target of mass $M_\mathrm{t}=10.5\,\mathrm{M}_{\oplus}$. Each different colour represents a different projectile mass ($M_\mathrm{p}$). Instead of a straight line (as in \citealt{Leinhardt2012}) our results appear to show two separate linear regimes: a shallower low energy one and a steeper high energy one.

\begin{figure}
\includegraphics[width=7.5cm]{{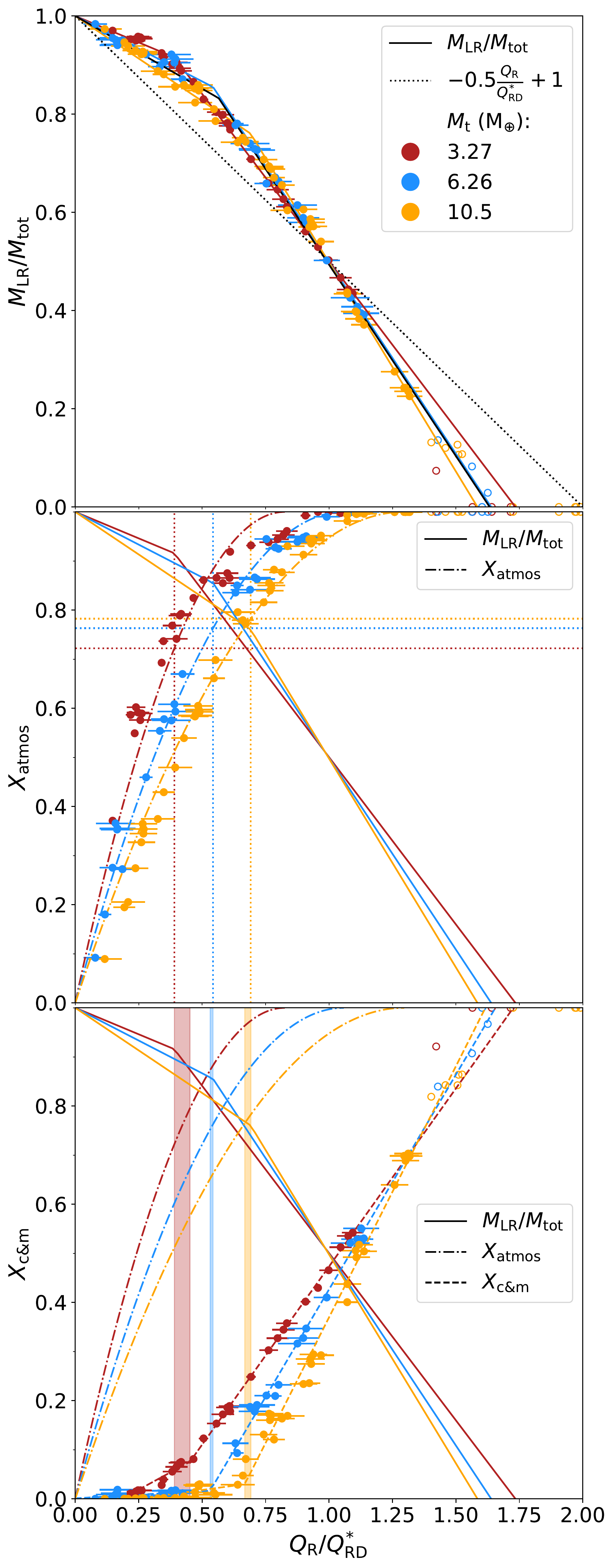}}
 \caption{\textit{Top:} $M_\mathrm{LR}$ versus $Q_\mathrm{R}/Q_{\mathrm{RD}}^*$ for all simulations. Colour indicates target mass, open circles represent points that were excluded from the fit. Coloured solid lines are best fits to data for a given $M_\mathrm{t}$. The black solid line is the best fit to the entire data set. The dotted black line is the universal law from \citet{Leinhardt2012}.
 \textit{Middle:} Fraction of atmosphere lost versus $Q_\mathrm{R}/Q_{\mathrm{RD}}^*$. The dashed coloured lines indicate best fits. The dotted vertical and horizontal lines show the specific energy of the break in the $M_{\mathrm{LR}}$ fit and the respective fraction of atmosphere loss.
 \textit{Bottom:} Fraction of core and mantle lost versus $Q_\mathrm{R}/Q_{\mathrm{RD}}^*$. Dashed lines are a broken linear fit. Shaded vertical sections show the difference between the specific energy of the break in the $M_{\mathrm{LR}}$ fit and the break in the core loss fit.
 For all 3 graphs the horizontal error bars represent the error in the $Q_{\mathrm{RD}}^*$ determination method.
 }
 
  \label{fig:QrMultipleTarget}
\end{figure}

While the existence of the atmosphere means that results split into two regimes, we test $Q_{\mathrm{RD}}^*$ to determine whether it still scales $Q_\mathrm{R}$ such that we obtain a mass independent fit (see Figure \ref{fig:QrMultipleTarget}). To determine $Q_{\mathrm{RD}}^*$, we fit our data with the following broken linear equation:

\begin{equation}
\label{eq:QRD_finder_fit}
\frac{M_{\mathrm{LR}}}{M_{\mathrm{tot}}} =
\begin{cases} 
      m_{\mathrm{1}}\left(Q_{\mathrm{R}}-Q_{\mathrm{piv}}\right)+c^{\mathrm{LR}} & Q_{\mathrm{R}}\leqslant Q_{\mathrm{piv}} \\
      m_{\mathrm{2}}\left(Q_{\mathrm{R}}-Q_{\mathrm{piv}}\right)+c^{\mathrm{LR}} & Q_{\mathrm{R}}>Q_{\mathrm{piv}},
   \end{cases}
\end{equation}
where $Q_{\mathrm{piv}}$ is the specific energy at the transition between the two regimes, measured to be the point where the two linear sections intersect, $c^{\mathrm{LR}}$ is the fraction of mass that has been lost at this transition energy, and $m_\mathrm{1}$ and $m_\mathrm{2}$ are the gradients of the two linear sections. Since jettisoning material should require input of energy, we fix this fit such that all mass is in the largest remnant for zero energy input. The gradient for the initial section thus becomes:
\begin{equation}
    m_\mathrm{1}=\frac{c^{\mathrm{LR}}-1}{Q_{\mathrm{piv}}}.
    \label{eq:m1_fixing}
\end{equation}

To ensure a robust fitting method, despite the low number of data points, we began by generating multiple random sets of input parameters for least squares fitting. We determined a best fitting set of output parameters for each input set, and chose the fit with the smallest squared residuals as our final best fit. The fits we obtain for each set of projectile and target masses can then be used to determine the catastrophic disruption threshold by measuring the energy at which half the mass of the system is in the largest remnant (as shown in Figure \ref{fig:QRNormalisation}). The errors on $Q_{\mathrm{RD}}^*$ were determined by propagating errors generated by the least squares fitting algorithm.

For solely rocky objects \citet{Leinhardt2012} showed that normalising the impact energy by the catastrophic disruption threshold will mean that mass fraction in the largest remnant will overlap for each set of different targets and projectiles (following equation \ref{eq:Leinhardtetal2012}). \citet{Marcus2010} showed that this relationship remains true for planets without atmospheres using similar SPH simulations with Gadget-2. The top panel of Figure \ref{fig:QrMultipleTarget} shows the mass of the largest remnant versus normalised specific impact energy for our simulations.  
The $Q_{\mathrm{RD}}^*$ scaling appears to be preserved for all projectile and target masses although the location of the pivot energy changes depending on the target. We fitted each of the targets individually in order to quantify this target dependence. 
The equation used for the fits is similar to that used for the un-normalised fits (equation \ref{eq:QRD_finder_fit}):
\begin{equation}
\label{eq:QRD_NormFit}
    \frac{M_{\mathrm{LR}}}{M_{\mathrm{tot}}} =
\begin{cases} 
      m_{\mathrm{1}}^{\mathrm{LR}}\left(\frac{Q_{\mathrm{R}}-Q_{\mathrm{piv}}}{Q_{\mathrm{RD}}^*}\right)+c^{\mathrm{LR}} & Q_{\mathrm{R}}\leqslant Q_{\mathrm{piv}}, \\
      m_{\mathrm{2}}^{\mathrm{LR}}\left(\frac{Q_{\mathrm{R}}-Q_{\mathrm{piv}}}{Q_{\mathrm{RD}}^*}\right)+c^{\mathrm{LR}} & Q_{\mathrm{R}}>Q_{\mathrm{piv}}
   \end{cases} 
\end{equation}
where we again assume that zero mass is lost for zero impact energy as in equation \ref{eq:m1_fixing}. The parameters obtained via these fits are found in Table \ref{tab:NormFits}. This table shows that there is a significant target mass dependence in the pivot energy, the pivot occurs at higher energies for higher target masses and atmosphere fractions.

 \begin{table}
  \caption{Parameters for fits using equation \ref{eq:QRD_NormFit} to the largest remnant mass with $Q_{\mathrm{RD}}^*$ normalised specific impact energy.}
  \label{tab:NormFits}
  \centering
  \begin{tabular}{ccccc}
    \hline
    M$_{\mathrm{t}}$ (M$_\oplus$) & m$_{\mathrm{2}}^{\mathrm{LR}}$  & c$^{\mathrm{LR}}$ & Q$_{\mathrm{piv}}$/Q$_{\mathrm{RD}}^{*}$ \\
    \hline
    3.27 & -0.68 $\pm$ 0.01 & 0.92 $\pm$ 0.01 & 0.39 $\pm$ 0.01\\ 
    6.26 & -0.78 $\pm$ 0.02 & 0.86 $\pm$ 0.01 & 0.54 $\pm$ 0.02\\ 
    10.5 & -0.85 $\pm$ 0.02 & 0.76 $\pm$ 0.01 & 0.69 $\pm$ 0.02\\ 
    \hline
  \end{tabular}
 \end{table}

  \begin{table}
  \caption{Parameters for fits to the fraction of atmosphere lost from the largest remnant with $Q_{\mathrm{RD}}^*$ normalised specific energy using equation \ref{eq:AtmosLoss_QRDNormalised}
  \label{tab:NormFits_Atmos}}
  \centering
  \begin{tabular}{ccccc}
    \hline
    Target Mass  (M$_\oplus$)& A \\
    \hline
    3.27 & 2.42 $\pm$ 0.06\\ 
    6.26 & 1.89 $\pm$ 0.03\\ 
    10.5 & 1.54 $\pm$ 0.02\\ 
    \hline
  \end{tabular}
 \end{table}

 \begin{table}
  \caption{Parameters for fits using equation \ref{eq:CoreLoss_QRDNormalised} to the fraction of core material lost from the largest remnant with $Q_{\mathrm{RD}}^*$ normalised specific energy.}
  \label{tab:NormFits_Core}
  \centering
  \begin{tabular}{ccccc}
    \hline
     M$_{\mathrm{t}}$ (M$_\oplus$) & m$_{\mathrm{1}}$ & m$_{\mathrm{2}}$  & c$^{\mathrm{c\&m}}$ & Q$_{\mathrm{piv}}$/Q$_{\mathrm{RD}}^{*}$ \\
    \hline
    3.27 & 0.28 $\pm$ 0.02 & 0.73 $\pm$ 0.01 & 0.08 $\pm$ 0.01 & 0.45 $\pm$ 0.01\\ 
    6.26 & 0.03 $\pm$ 0.03 & 0.88 $\pm$ 0.02 & 0.02 $\pm$ 0.01 & 0.53 $\pm$ 0.01\\ 
    10.5 & 0.07 $\pm$ 0.03 & 1.02 $\pm$ 0.02 & 0.03 $\pm$ 0.01 & 0.67 $\pm$ 0.01\\ 
    \hline
  \end{tabular}
 \end{table}

The middle panel of Figure \ref{fig:QrMultipleTarget} shows the fraction of atmosphere lost from the target versus specific impact energy normalised by the catastrophic disruption threshold. Our results show three separate curves, one for each target mass. The results show that less massive atmospheres of less massive planets are removed more easily than higher mass atmospheres of higher mass planets. 
We fit the atmosphere loss fraction, $X_{\mathrm{loss}}^{\mathrm{atmos}}$, for each target with a quadratic curve fixed to go through the origin and peak at total atmosphere removal:

\begin{equation}
\label{eq:AtmosLoss_QRDNormalised}
    X_{\mathrm{loss}}^{\mathrm{atmos}}=\frac{-A^2}{4}\left(\frac{Q_{\mathrm{R}}}{Q_{\mathrm{RD}}^*}\right)^{2}+A\frac{Q_{\mathrm{R}}}{Q_{\mathrm{RD}}^*},
\end{equation} 
where $A$ is the fit parameter. For results of these fits see Table \ref{tab:NormFits_Atmos}.
Some deviation from the quadratic fits is observed at higher energies, especially for the low mass target (3.27\,M$_\oplus$), this deviation may be due to resolution because there are fewer particles representing the atmosphere for lower mass targets.

We also investigate the dependence of core and mantle loss on the impact energy 
(bottom panel Figure \ref{fig:QrMultipleTarget}).
We observe a similar broken linear relation to the broken linear relation seen for the mass of the largest remnant (Figure \ref{fig:QRNormalisation} and Figure \ref{fig:QrMultipleTarget}, top). In this case we have a shallow gradient below the pivot energy (negligible for the two larger mass targets) indicating very little core and mantle mass loss, and a much steeper gradient (therefore much greater loss) above. For lower mass targets (lower atmosphere fractions) there is greater loss of core and mantle at low energies. 

In order to compare the energies of the pivots in the mass of the largest remnant and the core and mantle mass loss fraction, we fit the core and mantle loss fraction using the following equation:

\begin{equation}
\label{eq:CoreLoss_QRDNormalised}
     X_{\mathrm{loss}}^{\mathrm{c\&m}} =
\begin{cases} 
      m_{\mathrm{1}}^{\mathrm{c\&m}}\left(\frac{Q_{\mathrm{R}}-Q_{\mathrm{piv}}}{Q_{\mathrm{RD}}^*}\right)+c^{\mathrm{c\&m}} & Q_{\mathrm{R}}\leqslant Q_{\mathrm{piv}}, \\
      m_{\mathrm{2}}^{\mathrm{c\&m}}\left(\frac{Q_{\mathrm{R}}-Q_{\mathrm{piv}}}{Q_{\mathrm{RD}}^*}\right)+c^{\mathrm{c\&m}} & Q_{\mathrm{R}}>Q_{\mathrm{piv}}.
   \end{cases} 
\end{equation}

Parameters for these fits are given in Table \ref{tab:NormFits_Core}.
As can be seen from the coloured vertical bars in the bottom panel of Figure \ref{fig:QrMultipleTarget}, there appears to be a correlation between this pivot energy and the pivot energy found for the mass of the largest remnant. The similar energies of these pivot points implies that the break in slope for largest remnant mass is due to the break in slope for loss of core and mantle material. We therefore categorise impacts into two energy loss regimes: below the pivot energy we have atmosphere dominated loss, and above it we have substantial core and mantle loss.

An important point to note here is that at this pivot energy, the amount of atmosphere remaining is consistently $20-30\%$. This means that a single giant impact cannot remove all of a planet's atmosphere without also removing core and mantle material.

\subsection{Atmospheric Loss Scaling Law}

\begin{figure}
	\includegraphics[width=\columnwidth]{{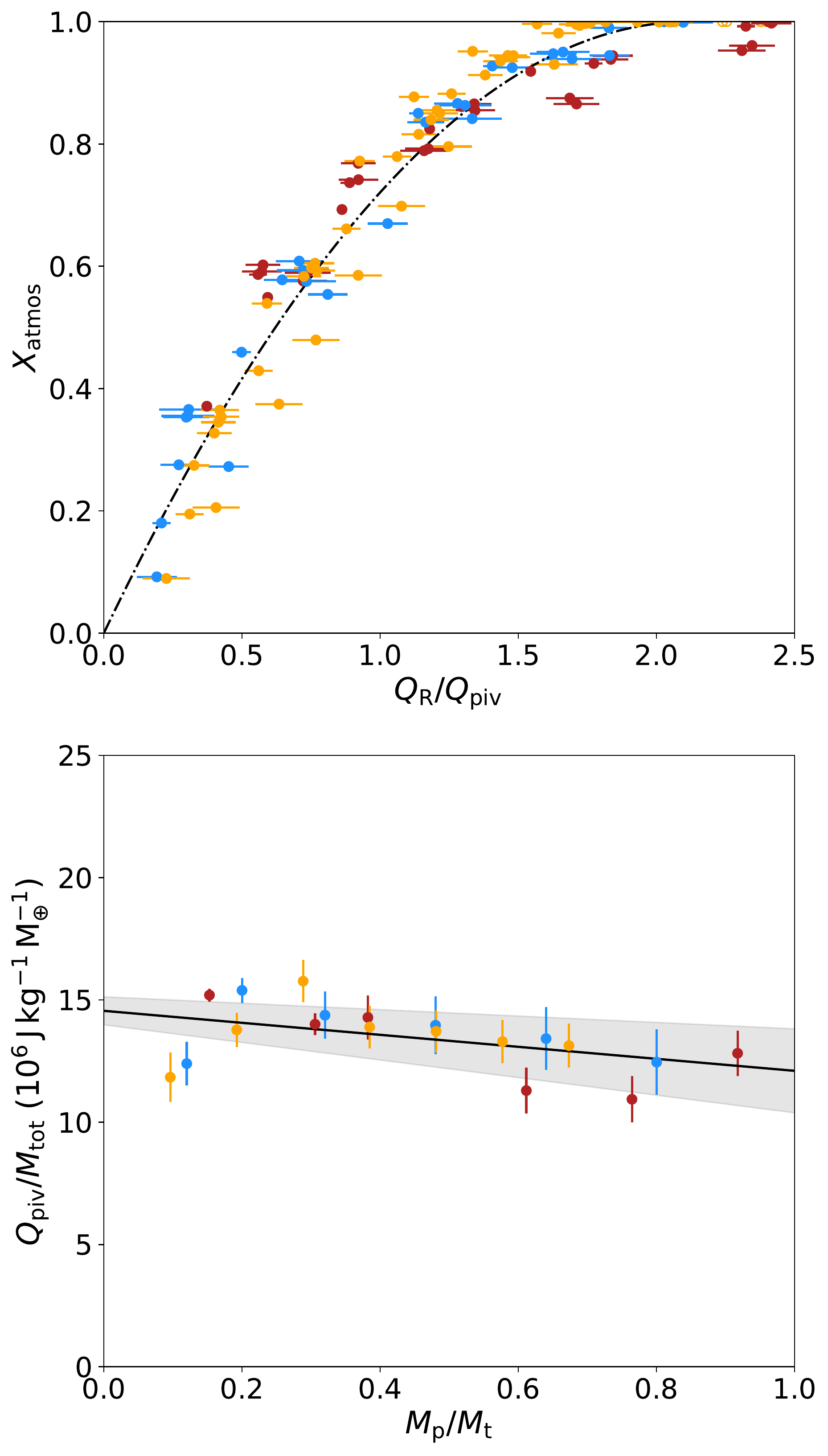}}
    \caption{
    \textit{Top:} Fractional atmosphere loss compared to specific impact energy scaled with respect to the pivot energy for each target mass. The amount of energy required to remove more atmosphere increases as the amount of atmosphere removed increases. Fitted to the data is a simple quadratic which has been fixed to cross the origin and have a peak at total atmosphere loss (equation \ref{eq:Xatm_fit}). \textit{Bottom:} Specific energy of the pivot normalised by the total mass of the system compared with projectile-target mass ratio. Mass normalised pivot energy appears to be approximately constant, decreasing slightly as projectile masses increase. The fit is given by equation \ref{eq:Qpiv_scaling}.
    }
    \label{fig:Qpiv}
\end{figure}

As we saw in Figure \ref{fig:QrMultipleTarget}, the atmosphere mass loss fraction coinciding with the pivot energy is consistently $70-80$ per cent. We therefore suggest a normalisation based on this pivot energy instead of $Q_{\mathrm{RD}}^*$. The results for this are shown in the top panel of Figure \ref{fig:Qpiv}. 
To build a scaling law for atmosphere loss, we have fitted the loss fraction using a quadratic curve (again with a constraint of zero mass loss at zero input energy and a peak at total atmosphere mass loss)

\begin{equation}
   X_{\mathrm{loss}}^{\mathrm{atmos}}=\frac{-A^2}{4}\left(\frac{Q_{\mathrm{R}}}{Q_{\mathrm{piv}}}\right)^{2}+A\frac{Q_{\mathrm{R}}}{Q_{\mathrm{piv}}},
   \label{eq:Xatm_fit}
\end{equation}

where our fitting parameter $A=0.94\pm0.01$. The other ingredient we require for building a scaling law for atmosphere loss is a way of predicting the pivot energy. 
After some experimentation we found a linear relation between the pivot energy divided by the total mass and the projectile target mass ratio for all atmosphere fractions tested (see bottom panel of Figure \ref{fig:Qpiv}):
\begin{equation}
    \frac{Q_{\mathrm{piv}}}{M_{\mathrm{tot}}}=-\left(2.5\pm1.1\right)\frac{M_{\mathrm{p}}}{M_{\mathrm{t}}}+\left(14.6\pm0.6\right)\;\; [10^6\,\mathrm{J\,kg^{-1}\,M_\oplus^{-1}}].
    \label{eq:Qpiv_scaling}
\end{equation} 

Equation \ref{eq:Qpiv_scaling} has quite a shallow gradient, meaning that the energy of the pivot is strongly dependent on the the total mass, i.e. the impact energy is strongly dependent on the square of the total mass, implying it is related to the total gravitational binding energy of the system. The error envelope in Figure \ref{fig:Qpiv} is generated from the variance of the fit parameters. Together Equations \ref{eq:Xatm_fit} and \ref{eq:Qpiv_scaling} can be used to predict the mass fraction of atmosphere lost from the target for any head-on giant impact in the regime tested.

\section{Discussion}

\subsection{Caveats}

\label{sec:caveats}

\subsubsection{Atmosphere fraction effects}
We chose the initial atmosphere fraction for each target planet based on the most likely atmosphere fractions from the Bern global planet formation model (equation \ref{eq:atmos_frac_pred}). One issue with this method is that it makes distinguishing between the effects of target mass and atmosphere fraction somewhat difficult, as both are directly related. 
An important result that requires further investigation is the relation between atmosphere fraction and pivot energy. One might expect that less massive atmospheres would require less energy to remove and thus core and mantle material would begin to be removed earlier, decreasing the pivot energy. We observe a decrease in pivot energy but we cannot distinguish whether this is due to a smaller atmosphere fraction or a shallower gravitational potential well from a less massive target. In the future we could test the cause of the decrease in pivot energy by simulating collisions involving targets of the same masses with different atmosphere fractions. 

\subsubsection{Equation of State effects}
One simplification we have made in this work is to use ideal gas atmospheres as opposed to a more realistic equation of state. For Earth atmospheric pressures this would have a negligible effect, but we are dealing with significantly more massive atmospheres (roughly seven orders of magnitude). For such massive atmospheres the pressures and densities at the base of the atmosphere are such that the assumption that molecules do not interact (inherent in using an ideal gas) may not be realistic. The lack of inter-particle interactions means that we have unrealistically high densities at high pressures, (the base of our atmospheres post equilibration being $1.2$--$19.2\times10^5$ atm or $1.2$--$19.2\times10^9$ $\mathrm{GPa}$). One effect of this is a significant compression of the atmosphere when equilibrating a Gadget-2 planet. During the equilibration the radii of atmospheres shrank on average by a factor of two. Despite these issues, we have used an ideal gas as a starting point, more realistic equations of state should be the subject of a future study.

\subsubsection{Impact Angle effects}
For impact angle we have also elected to investigate the simplest case -- head-on collisions. In general, we would expect an average collision angle of $45^{\circ}$ \citep{Shoemaker1962StratigraphicScale}. \citet{Leinhardt2012} present in their prescription a method to relate the mass loss of an off angle collision with a head-on one, which uses an interacting mass. In the impact scenarios considered in this work we would expect the impact angle correction to become more complicated due to the density contrast between atmosphere and mantle. \citet{Hwang2017b} and \citet{Hwang2017a} have previously done work on grazing collisions where the cores and mantles did not touch. Their results show that mass loss follows a power-law dependence on impact parameter. We plan to investigate this further in future work.

\subsection{Catastrophic Disruption threshold Scaling}

\begin{figure}
    \centering
    \includegraphics[width=\columnwidth]{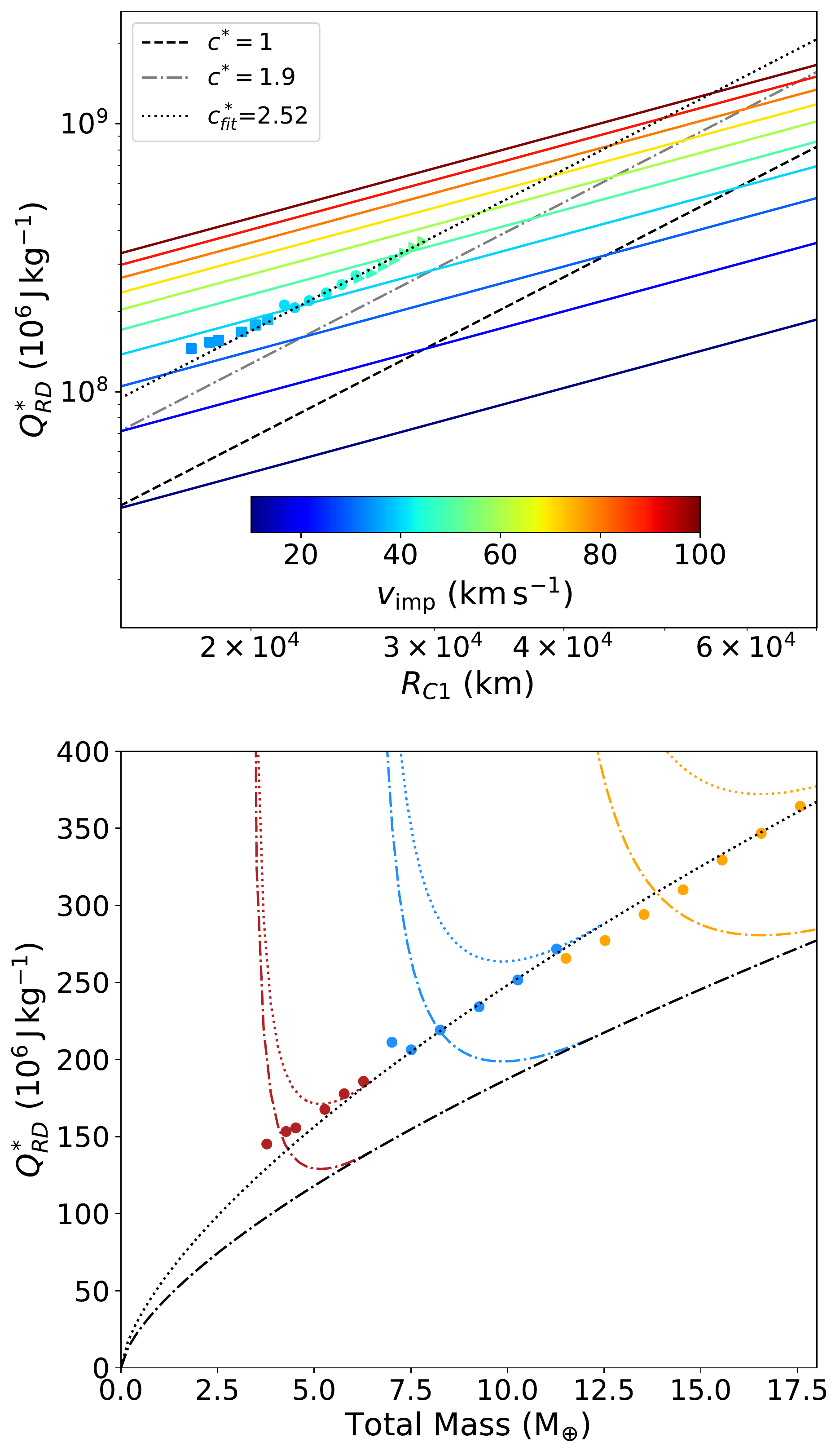}
    \caption{
    \textit{Top:} 
    Catastrophic disruption threshold $Q_{\mathrm{RD}}^{*}$ compared with $R_{\mathrm{C1}}$ the radius a spherical body would have if it had the total system mass and a density of 1000\,$\mathrm{kg\,m^{-3}}$, with black and grey lines showing their predicted relationship for different values of the strength parameter $c^*$ following equation \ref{eq:QRD2012Scaling}. We observe a value of $c^*=2.52$, which is a $32\%$ increase compared to \citet{Leinhardt2012}'s value of $1.9$ for solely rocky bodies. Coloured lines show the predicted catastrophic disruption threshold for equal mass collisions for particular velocities. The colour of each data point indicates the relative velocity the pair of planets would have for each collision if they were equal mass.  
    \textit{Bottom:} Comparison between the catastrophic disruption threshold, and the total mass of each set of collisions. Black lines detail predictions of the catastrophic disruption threshold for each total mass for pure energy scaling for different values of $c^{*}$, dotted is our measured value of $2.52$, dot-dashed is \citet{Leinhardt2012}'s value of $1.9$, coloured lines show predictions for pure momentum scaling. Our results (dots with colours representing target mass as per figure \ref{fig:QrMultipleTarget}) appear to follow the prediction for pure energy scaling.
    }
    \label{fig:QRD*_Rc1}
\end{figure}

\citet{Leinhardt2012} predict that for equal mass impacts the catastrophic disruption threshold should obey:
\begin{equation}
Q_{\mathrm{RD}(\gamma=1)}^{*}=c^*\frac{4}{5}\pi\rho_{\mathrm{1}}GR_{\mathrm{C1}}^2
\label{eq:QRD2012Scaling}
\end{equation}
where $\gamma = M_\mathrm{p}/M_\mathrm{t}$, $c^*$ is a measure of the catastrophic disruption threshold in units of the gravitational binding energy (measured to be $c^*=1.9 \pm 0.3$ for hydrodynamical simulations of large rocky planets),  $\rho_{\mathrm{1}}=1000\,\mathrm{kg\,m^{-3}}$, and $R_{\mathrm{C1}} = \left(\frac{3M_{\mathrm{tot}}}{4\pi\rho_{\mathrm{1}}}\right)^{\frac{1}{3}}$ is the radius a spherical body would have if it had the total system mass and a density of $\rho_{\mathrm{1}}$. 

\citet{Leinhardt2012} also propose a further correction for this catastrophic disruption threshold for collisions of different mass projectiles and targets,
\begin{equation}
    Q_{\mathrm{RD}}^{*}=Q_{\mathrm{RD}(\gamma=1)}^{*}\left(\frac{(\gamma+1)^2}{4\gamma}\right)^{\frac{2}{3\bar{\mu}}-1},
\end{equation}
which is dependent upon a parameter $\bar{\mu}$ where $\bar{\mu}$ has values between $1/3$ (pure momentum scaling) and $2/3$ (pure energy scaling). This multiplicative correction becomes unity for all mass ratios with perfect energy scaling because the index becomes zero.

Figure \ref{fig:QRD*_Rc1} details how our results compare to \citeauthor{Leinhardt2012}'s predictions. Our results run parallel to the prediction of equation \ref{eq:QRD2012Scaling} indicating a larger $c^*$ value of 2.52 (compared to the value given in \citealt{Leinhardt2012}). The larger $c^*$ means that a greater excess over the binding energy is required to remove material from planets with atmospheres than those without. This is presumably due to the increased compressibility and decreased viscosity of atmosphere compared to mantle and core material.  

It should be noted that we appear to be observing pure energy scaling for these collisions as no mass ratio correction is required, this is illustrated best in the bottom part of Figure \ref{fig:QRD*_Rc1}. Here, black lines represent the predicted catastrophic disruption threshold using energy scaling, whilst coloured lines represent the momentum scaling which \citet{Leinhardt2012} predicts we should be close to. As can be observed, our results follow the energy scaled prediction closely. This runs contrary to the results in \cite{Leinhardt2012} which predicts near perfect momentum scaling. Presumably this difference is due to the presence of the atmosphere, as the \citet{Leinhardt2012} method is based on \citet{Holsapple1987PointMechanics.}'s crater scaling, where they show that porous materials tend to follow momentum scaling, while perfect gases tend to follow energy scaling.

\subsection{Mass loss efficiency comparisons}

\begin{figure}
    \centering
    \includegraphics[width=\columnwidth]{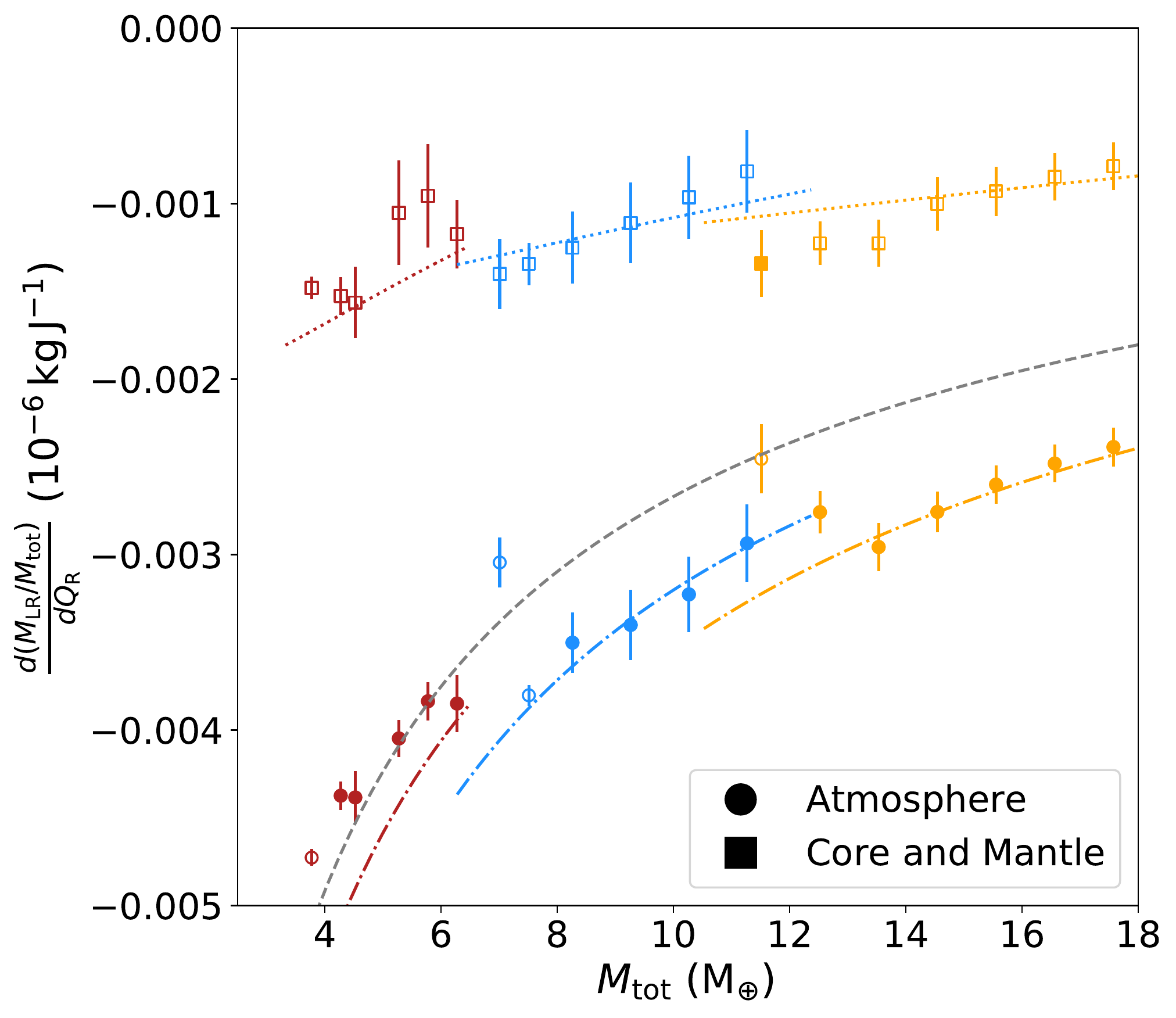}
    \caption{Comparisons between the mass loss efficiency for core and mantle, and for atmosphere to the total mass of the system. Atmosphere loss shows a decreased efficiency in comparison to core and mantle loss, as might be expected considering the increased compressibilty of atmosphere material. Filled shapes are gradients where we obtained 4 or greater data points for that line, open are where we have 3 or fewer. The grey line shows the loss efficiency predicted by \citet{Leinhardt2012} for rocky material, the coloured line beneath them is this value multiplied by 1 plus the atmosphere fraction for that particular target as per equation \ref{eq:CoreGrad_mod}. This appears to show reasonable correlation with our results especially for higher mass targets. The coloured dotted line above this shows a prediction of what the atmosphere gradient must be that uses all previously derived scaling laws, is linear and passes through zero mass loss for zero input energy (equation \ref{eq:AtmosGrad}). Our results show some degree of correlation with this value, but we do not have a high enough density of data in this region to effectively probe the accuracy of this prediction. }
    \label{fig:Gradient_scaling}
\end{figure}

\citet{Leinhardt2012} predict that the mass of the largest post-collision remnant scales with normalised specific impact energy (equation \ref{eq:Leinhardtetal2012}). This equation gives a gradient with respect to $Q_\mathrm{R}$ for the largest remnant mass of
\begin{equation} 
   \frac{d\left(M_{\mathrm{LR}}/M_{\mathrm{tot}}\right)}{dQ_{\mathrm{R}}}=-\frac{0.5}{Q_{\mathrm{RD}}^{*(\mathrm{LS12})}}.
    \label{eq:CoreGrad_LS12}
\end{equation}
In this work we found that the largest remnant mass as a function of specific energy falls into two regimes (see Figure \ref{fig:QRNormalisation}) with two different gradients. 
Figure \ref{fig:Gradient_scaling} compares the gradients measured from our results to \citeauthor{Leinhardt2012}'s prediction.

We can consider these gradients as a measurements of the efficiency of mass loss in each regime. The efficiency of mass loss for the core and mantle loss region is typically at least double that for the atmosphere dominated loss regime, this is presumably due to the increased compressibility of the atmosphere. 

As can be observed, the \citet{Leinhardt2012} model (equation \ref{eq:CoreGrad_LS12} above) appears to under predict the efficiency of mantle and core material loss for planets with substantial envelopes. This is likely due to the pressure of the atmosphere above providing a resisting force to reduce mantle loss. Once the atmosphere is removed the specific energy of the impact is higher than would have been necessary to remove significant mantle if no atmosphere was present, so more material is being removed per unit of specific energy. 

We have found that an approximate prediction of the gradient in the core and mantle loss regime is given by:
\begin{equation}
    \frac{d\left(M_{\mathrm{LR}}/M_{\mathrm{tot}}\right)}{dQ_{\mathrm{R}}}=-\frac{0.5(1+f_{\mathrm{atmos}})}{Q_{\mathrm{RD}}^{*(\mathrm{LS12})}},
    \label{eq:CoreGrad_mod}
\end{equation}
where $f_{\mathrm{atmos}}$ is the atmosphere fraction of the target. For our higher mass targets with more massive atmospheres, this seems to be a reasonably good predictor of loss efficiency; for the smallest target with the lowest mass atmosphere, however, the initial prediction from equation \ref{eq:CoreGrad_LS12}  seems to match more closely.

To predict the loss efficiency (gradient) in the atmosphere loss dominated regime, we assume that there is zero mass loss for zero input energy, and use our scaling laws for the pivot energy (equation \ref{eq:Qpiv_scaling}) and catastrophic disruption threshold (equation \ref{eq:QRD2012Scaling}) along with our correction to the gradient in the core and mantle loss regime (equation \ref{eq:CoreGrad_mod}) to predict the fraction of material in the largest remnant at the pivot energy. We also assume that the gradient in the atmosphere loss regime is constant with respect to energy (as in Equations \ref{eq:QRD_finder_fit} and \ref{eq:m1_fixing}). Combining these equations we obtain the following relation,
\begin{equation}
    \frac{d\left(M_{\mathrm{LR}}/M_{\mathrm{tot}}\right)}{dQ_{\mathrm{R}}}=\frac{-0.5-m^{\mathrm{LR}}_{\mathrm{2}}(Q_{\mathrm{RD}}^{*(\mathrm{New})}-Q_{\mathrm{piv}})}{Q_{\mathrm{piv}}}.
    \label{eq:AtmosGrad}
\end{equation}
Our results show a reasonable match with this prediction for the gradient in the atmosphere loss regime. However, we do not have a sufficient data in this region to fully probe the accuracy of this scaling; for example we cannot test whether atmosphere loss begins at zero impact energy (as we have assumed in deriving the scaling) or requires some small initial energy input.

\subsection{Implications}

\begin{figure}
    \centering
    \includegraphics[width=\columnwidth]{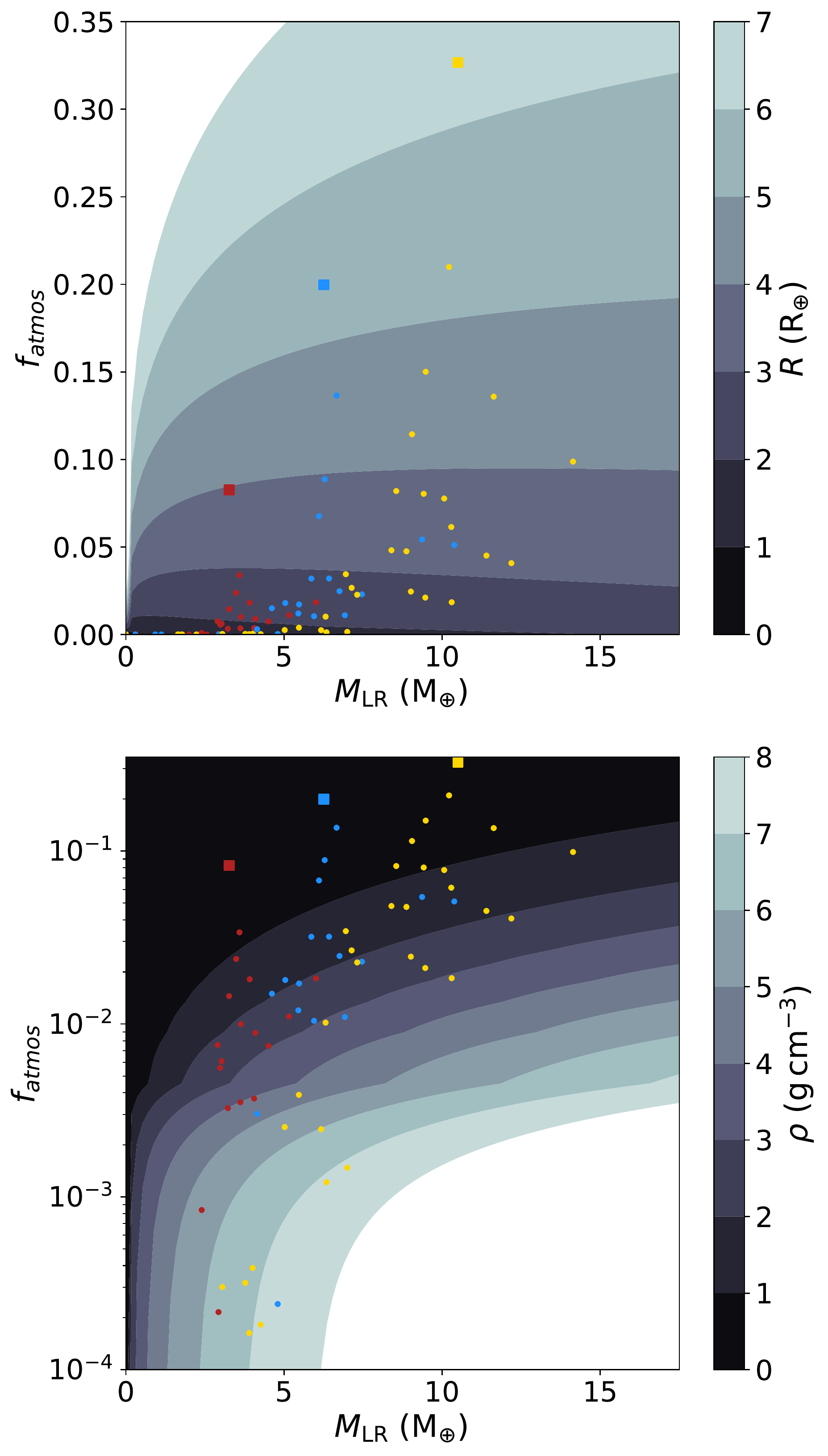}
    \caption{\textit{Top:} Radius as a function of mass following the prescription of \citet{Lopez2014}, compared against our initial targets (squares) and also our post-collision largest remnants (circles), the same colour scheme for target masses is used as for the rest of the paper.
    \textit{Bottom:} Density as a function of mass calculated using the radius above, this is similarly compared against our initial targets (squares) and also our post-collision largest remnants (circles), collisions with full atmosphere removal have been removed so atmosphere fraction could be plotted logarithmically.}
    \label{fig:Density_Predictions}
\end{figure}

This study was motivated by the density disparity observed in exoplanet systems. The post-impact bodies in our simulations are hot and inflated, often with a large mass of vaporised silicate, and thus do not represent the structure of planets millions of years after their final giant impacts  \citep{Lock2017,Carter2020,Lock2020TheImpact}. To determine the bulk density of the remnants from their post-collision material composition, we use the approximation suggested in \citet{Lopez2014}. In their model the total radius of a planet, $R_{\mathrm{planet}}$, from which we determine the density, can be calculated from summing the radial contributions of the following three components:
the core and mantle, which has a power law relation to mass;  
the convective envelope, which is dependent on the temperature of the atmosphere, which itself is a function of stellar flux and planet age; 
and the radiative atmosphere (also dependent on temperature). 
Here, we ignore the contribution of the radiative atmosphere due to its small effect ($\sim0.1\mathrm{R_{\oplus}}$) 
We used an age of $5\,\mathrm{Gyr}$ for our comparison planets (see Figure \ref{fig:Density_Predictions}) as this is the most common age for stars in the local galactic neighbourhood. We  used a flux, $F_{\mathrm{planet}}$, of $100\,\mathrm{F_{\oplus}}$ as the type of planets we simulated are most commonly observed at $\sim0.1$\,au around Sun-like stars.
Figure \ref{fig:Density_Predictions} shows contour plots of the radius (top) and density (bottom) as a function of mass and atmosphere fraction according to the \citet{Lopez2014} approximation described above. 

The prediction we obtained for the envelope radius of our initial targets using the \citet{Lopez2014} model was within 10 per cent of the radius of our initial thermodynamic profiles. This radius was, however, significantly larger than our Gadget-2 targets due to the compression caused by using an ideal gas. We also note that we could not reach the 20\,mBar pressures which \citet{Lopez2014} consider to be the edge of the atmosphere with computationally practical resolutions. 

For our collisions, we always observe a decrease in atmosphere fraction. 
This decrease in atmosphere fraction means that for all the collisions we simulate, except for those resulting in largest remnants below the resolution limit, we observe an increase in density as shown in the bottom panel of Figure \ref{fig:Density_Predictions}.

\subsection{New prescriptions for atmosphere loss and largest remnant mass}

Here we summarise the process one would need to use to predict the atmosphere loss from any giant impact in the regime probed by this paper, as well as the modifications to the \citet{Leinhardt2012} prescription for largest post-collision remnant mass for an arbitrary head on collision between a Mini-Neptune with a significant gaseous envelope and a lower mass Super-Earth without an atmosphere. This algorithm can be incorporated into $N$-body codes and population synthesis models.

\begin{enumerate}
    \item \noindent For a given collision scenario ($M_\mathrm{p}$, $M_\mathrm{t}$ and $V_\mathrm{imp}$), calculate the specific relative kinetic energy of the impact,
    \begin{equation}
        Q_\mathrm{R}=\frac{1}{2}\mu\frac{V_\mathrm{imp}^2}{M_{\mathrm{tot}}},
    \end{equation}
    where $M_{\mathrm{tot}}=M_\mathrm{t}+M_\mathrm{p}$, and $\mu = M_\mathrm{t}M_\mathrm{p}/(M_\mathrm{t}+M_\mathrm{p})$ is the reduced mass, $M_\mathrm{t}$ is the target mass, $M_\mathrm{p}$ the projectile mass and $V_{\mathrm{imp}}$ is the impact velocity.
    
    \item \noindent Then, calculate the specific kinetic energy of the transition between the atmosphere loss and core and mantle loss regimes,
    \begin{equation}
        Q_{\mathrm{piv}}=\left(\frac{M_{\mathrm{tot}}}{M_\oplus}\right)\left(-2.45\frac{M_\mathrm{p}}{M_\mathrm{t}}+14.56\right) \;\; [10^6\,\mathrm{J\,kg^{-1}}].
    \end{equation}
    
    \item \noindent Calculate the catastrophic disruption threshold,
    \begin{equation}
    Q_{\mathrm{RD}}^{*}=c^*\frac{4}{5}\pi\rho_{\mathrm{1}}GR_{\mathrm{C1}}^2,
    \end{equation}
    using a value of $c^*=2.52$ for collisions involving planets with atmospheres ($Q_{\mathrm{RD}}^{*(\mathrm{New})}$) and $c^*=1.9$ ($Q_{\mathrm{RD}}^{*(\mathrm{LS12})}$, as in \citealt{Leinhardt2012})  for targets with no atmosphere. $\rho_{\mathrm{1}}=1000\,\mathrm{kg\,m^{-3}}$, and $R_{\mathrm{C1}} = \left(\frac{3M_{\mathrm{tot}}}{4\pi\rho_{\mathrm{1}}}\right)^{\frac{1}{3}}$ is the radius a spherical body would have if it had the total system mass and a density of $\rho_{\mathrm{1}}$.\\
    
    \item \noindent Calculate the gradients of each of the linear sections of the largest remnant mass fraction relation. For the core and mantle loss regime this is
    \begin{equation}
        m_{\mathrm{c\&m}}=\frac{-0.5(1+f_{\mathrm{atmos}})}{Q_{\mathrm{RD}}^{*(\mathrm{LS12})}}
    \end{equation} 
    where $f_{\mathrm{atmos}}$ is the mass fraction of the target which is atmosphere.

    \item \noindent Zero impact energy means zero mass loss, therefore the gradient for the atmosphere loss dominated part of the relation is, from equation \ref{eq:m1_fixing},
    \begin{equation}
        m_{\mathrm{atmos}}=\frac{m_{\mathrm{c\&m}}(Q_{\mathrm{piv}}-Q_{\mathrm{RD}}^{*(\mathrm{New})})-0.5}{Q_{\mathrm{piv}}}.
    \end{equation}
    
    \item \noindent Next, calculate the super-catastrophic disruption threshold, taking this to be where <10 per cent of the initial mass ends up in the largest remnant (following \cite{Leinhardt2012}) we obtain
    \begin{equation}
        Q_{\mathrm{supercat}}=Q_{\mathrm{RD}}^{*(\mathrm{New})}-\frac{0.4}{m_{\mathrm{c\&m}}}.
    \end{equation}
    
    \item \noindent Then, the total mass in the largest remnant is:
    \begin{equation}
        \frac{M_{\mathrm{LR}}}{M_{\mathrm{tot}}}=
        \begin{cases}
             m_{\mathrm{atmos}}Q_{\mathrm{R}}+1 & 0<Q_{\mathrm{R}}<Q_{\mathrm{piv}} \\ \\
             m_{\mathrm{c\&m}}(Q_{\mathrm{R}}-Q_{\mathrm{RD}}^{*(\mathrm{New})})+0.5 & Q_{\mathrm{piv}}<Q_{\mathrm{R}}<Q_{\mathrm{supercat}}.\\

        \end{cases}
    \end{equation}
    We did not probe the super-catastrophic disruption regime in this study due as this would require much higher resolution; for collisions in this energy regime we recommend using the prescription of \cite{Leinhardt2012}.\\
    
    \item \noindent Finally, the atmosphere fraction lost is,

    \begin{equation}
    \chi_{\mathrm{loss}}^{\mathrm{atmos}} =
    \begin{cases} 
          \frac{-0.94^2}{4}\left(\frac{Q_{\mathrm{R}}}{Q_{\mathrm{piv}}}\right)^{2}+0.94\frac{Q_{\mathrm{R}}}{Q_{\mathrm{piv}}} & \frac{Q_{\mathrm{R}}}{Q_{\mathrm{piv}}}<2.12 \\
          1 & \frac{Q_{\mathrm{R}}}{Q_{\mathrm{piv}}}>2.12.
    \end{cases}
    \end{equation}

\end{enumerate}

\section{Summary}

In this paper we present the results from a series of SPH simulations of head-on collisions of planets in which the target has a significant atmosphere. Our findings are summarised below:
\begin{itemize}
    \item Giant impacts can have sufficient energy to remove large fractions of mass from the target planet; the mass lost is dependent upon the specific kinetic energy of the impact.
    \item Giant impacts can result in substantial increases in the densities of Mini-Neptune planets by ejecting a fraction of their atmospheres.
    \item The fraction of mass lost splits into two regimes -- at low specific impact energies only the outer layers are ejected corresponding to atmosphere dominated loss, at higher energies material deeper in the potential is excavated resulting in significant core and mantle loss.
    \item Approximately twenty per cent of the initial atmosphere remains at the transition between the two regimes.
    \item A single collision cannot remove all the atmosphere without also removing a significant percentage of mantle material. 
    \item Mass removal is less efficient in the atmosphere loss dominated regime compared to the core and mantle loss regime.
    \item The specific energy of this transition (pivot energy, $Q_{\mathrm{piv}}$) scales linearly with the ratio of projectile to target mass for all projectile-target mass ratios measured:\\ $Q_{\mathrm{piv}}=M_{\mathrm{tot}}\left(-2.45\frac{M_\mathrm{p}}{M_\mathrm{t}}+14.56\right) \;\;[10^6\,\mathrm{J\,kg^{-1}\,M_\oplus^{-1}}]$.
    \item The fraction of atmosphere lost is well approximated by a quadratic in terms of the ratio of specific energy to transition energy: $X_{\mathrm{loss}}^{\mathrm{atmos}} = \frac{-0.94^2}{4}\left(\frac{Q_{\mathrm{R}}}{Q_{\mathrm{piv}}}\right)^{2}+0.94\frac{Q_{\mathrm{R}}}{Q_{\mathrm{piv}}}$, for $Q_{\mathrm{R}}<2.12\,Q_{\mathrm{piv}}$, and total atmosphere loss for energies greater than this.
\end{itemize}

\section*{Acknowledgements}
This work was carried out using the computational facilities of the Advanced Computing Research Centre, University of Bristol - http://www.bristol.ac.uk/acrc/. 
TD acknowledges support from an STFC studentship. PJC acknowledges support from UC Office of the President grant LFR-17-449059.
CM acknowledges support from the Swiss National Science Foundation under grant BSSGI0$\_$155816 ``PlanetsInTime''. Parts of this work have been carried out within the framework of the NCCR PlanetS supported by the Swiss National Science Foundation. 
This research has made use of the NASA Exoplanet Archive, which is operated by the California Institute of Technology, under contract with the National Aeronautics and Space Administration under the Exoplanet Exploration Program.



\bibliographystyle{mnras}
\bibliography{library}



\appendix

\section{Further Impacts}

We present results in this paper for collisions against 3 separate simulated Super-Earth targets, a $3.27\,\mathrm{M}_{\oplus}$ target with a $0.27\,\mathrm{M_{\oplus}}$ atmosphere, $1\,\mathrm{M_{\oplus}}$ core and $2\,\mathrm{M_{\oplus}}$ mantle  (initial conditions and results given in Table \ref{tab:3_0.27}); a $6.26\,\mathrm{M}_{\oplus}$ one with a $1.25\,\mathrm{M_{\oplus}}$ atmosphere, $1.67\,\mathrm{M_{\oplus}}$ core and $3.34\,\mathrm{M_{\oplus}}$ mantle (Table \ref{tab:5_1.25}); and a $10.5\,\mathrm{M}_{\oplus}$ one with a $3.43\,\mathrm{M_{\oplus}}$ atmosphere, $2.36\,\mathrm{M_{\oplus}}$ core and $4.71\,\mathrm{M_{\oplus}}$ mantle (Table \ref{tab:7_3.43}).

\begin{table*}
\caption{
Parameters for head-on collisions between a $3.27\,\mathrm{M}_{\oplus}$ target ($M_\mathrm{t}^{\mathrm{core}} = 3.0 \, \mathrm{M}_{\oplus}$, $M_\mathrm{t}^{\mathrm{atmos}} = 0.27 \, \mathrm{M}_{\oplus}$) with a mantle surface radius of $1.31\,\mathrm{R}_{\oplus}$, and an atmosphere scale height of $0.52\,\mathrm{R}_{\oplus}$, and projectiles of mass $M_{\mathrm{p}}$. Similar to Tables \ref{tab:5_1.25} and \ref{tab:7_3.43}.
}
\label{tab:3_0.27}
\begin{tabular}{lccccccccccccccccc}
\hline
ID&$M_{\mathrm{p}}$ &$R_{p}$ &$v_{\mathrm{imp}}^{\mathrm{pred}}$ &$\frac{v_{\mathrm{imp}}^{\mathrm{pred}}}{v_{\mathrm{esc}}}$ &$v_{\mathrm{imp}}^{\mathrm{meas}}$ &$\frac{v_{\mathrm{imp}}^{\mathrm{meas}}}{v_{\mathrm{esc}}}$ &$v_{\mathrm{init}}$ &$\frac{v_{\mathrm{init}}}{v_{\mathrm{esc}}}$ &$S$ &$M_{\mathrm{LR}}$ &$M_{\mathrm{LR}}^{\mathrm{atmos}}$ &$M_{\mathrm{LR}}^{\mathrm{core}}$ &$\frac{M_{\mathrm{LR}}}{M_{\mathrm{tot}}}$ & $X_{\mathrm{loss}}^{\mathrm{atmos}}$ & $X_{\mathrm{loss}}^{\mathrm{c\&m}}$ & Category \\
&$\mathrm{M}_{\oplus}$ &$\mathrm{R}_{\oplus}$ &$\mathrm{km}\,\mathrm{s}^{-1}$ & &$\mathrm{km}\,\mathrm{s}^{-1}$ & &$\mathrm{km}\,\mathrm{s}^{-1}$ & &$\mathrm{R}_{\oplus}$ &$\mathrm{M}_{\oplus}$ &$\mathrm{M}_{\oplus}$ &$\mathrm{M}_{\oplus}$& & & & \\

\hline
3-0& 0.50& 0.78& 20.00& 1.46& 19.28& 1.41& 14.98& 1.10& 9.59& 3.66& 0.17& 3.49& 0.97& 0.37& 0.0& AL-CM\\
3-1& 0.50& 0.78& 25.00& 1.83& 24.32& 1.78& 21.20& 1.55& 9.59& 3.59& 0.12& 3.47& 0.95& 0.56& 0.01& AL-CM\\
3-2& 0.50& 0.78& 30.00& 2.19& 29.31& 2.14& 26.92& 1.97& 9.59& 3.49& 0.08& 3.41& 0.92& 0.7& 0.03& AL-CM\\
3-3& 0.50& 0.78& 35.00& 2.56& 34.29& 2.51& 32.39& 2.37& 9.60& 3.27& 0.05& 3.22& 0.87& 0.81& 0.08& AL-CA\\
3-4& 0.50& 0.78& 40.00& 2.92& 39.25& 2.87& 37.74& 2.76& 9.59& 2.90& 0.02& 2.88& 0.77& 0.93& 0.18& AL-CE\\
3-5& 0.50& 0.78& 50.00& 3.66& 49.14& 3.59& 48.21& 3.53& 9.59& 2.00& 0.00& 2.00& 0.53& 1.0& 0.43& TAL-CE\\
3-6& 1.00& 0.97& 20.00& 1.40& 19.30& 1.35& 14.77& 1.04& 10.13& 4.07& 0.11& 3.96& 0.95& 0.59& 0.01& AL-CM\\
3-7& 1.00& 0.97& 25.00& 1.75& 24.37& 1.71& 21.05& 1.48& 10.12& 3.92& 0.07& 3.85& 0.92& 0.74& 0.04& AL-CM\\
3-8& 1.00& 0.97& 30.00& 2.10& 29.40& 2.06& 26.80& 1.88& 10.14& 3.55& 0.04& 3.51& 0.83& 0.85& 0.12& AL-CA\\
3-9& 1.00& 0.97& 35.00& 2.46& 34.41& 2.41& 32.30& 2.27& 10.13& 3.03& 0.02& 3.01& 0.71& 0.93& 0.25& AL-CA\\
3-10& 1.00& 0.97& 40.00& 2.81& 39.39& 2.76& 37.66& 2.64& 10.12& 2.40& 0.00& 2.40& 0.56& 1.0& 0.4& TAL-CE\\
3-11& 1.00& 0.97& 50.00& 3.51& 49.31& 3.46& 48.15& 3.38& 10.14& 0.32& 0.00& 0.32& 0.07& 1.0& 0.92& SCD\\
3-12& 1.25& 1.03& 20.00& 1.38& 19.29& 1.33& 14.60& 1.00& 10.31& 4.30& 0.11& 4.19& 0.95& 0.59& 0.02& AL-CM\\
3-13& 1.25& 1.03& 25.00& 1.72& 24.39& 1.68& 20.93& 1.44& 10.31& 4.08& 0.06& 4.02& 0.9& 0.78& 0.06& AL-CA\\
3-14& 1.25& 1.03& 30.00& 2.06& 29.43& 2.03& 26.70& 1.84& 10.31& 3.64& 0.04& 3.60& 0.8& 0.85& 0.15& AL-CA\\
3-15& 1.25& 1.03& 35.00& 2.41& 34.43& 2.37& 32.22& 2.22& 10.31& 2.98& 0.02& 2.97& 0.66& 0.93& 0.3& AL-CE\\
3-16& 1.25& 1.03& 40.00& 2.75& 39.42& 2.71& 37.59& 2.59& 10.32& 2.27& 0.00& 2.27& 0.5& 1.0& 0.47& TAL-CE\\
3-17& 1.25& 1.03& 50.00& 3.44& 49.35& 3.40& 48.09& 3.31& 10.31& -& -& -& -& 1.0& -& SCD\\
3-18& 2.00& 1.18& 20.00& 1.31& 19.32& 1.26& 14.03& 0.92& 10.77& 5.03& 0.11& 4.92& 0.95& 0.59& 0.02& AL-CM\\
3-19& 2.00& 1.18& 25.00& 1.63& 24.38& 1.59& 20.54& 1.34& 10.78& 4.69& 0.06& 4.63& 0.89& 0.78& 0.07& AL-CA\\
3-20& 2.00& 1.18& 30.00& 1.96& 29.42& 1.92& 26.39& 1.73& 10.79& 4.10& 0.04& 4.06& 0.78& 0.85& 0.19& AL-CA\\
3-21& 2.00& 1.18& 35.00& 2.29& 34.46& 2.25& 31.96& 2.09& 10.78& 3.23& 0.01& 3.21& 0.61& 0.96& 0.36& TAL-CA\\
3-22& 2.00& 1.18& 40.00& 2.62& 39.46& 2.58& 37.37& 2.44& 10.78& 2.29& 0.00& 2.29& 0.43& 1.0& 0.54& TAL-CE\\
3-23& 2.00& 1.18& 50.00& 3.27& 49.42& 3.23& 47.92& 3.13& 10.79& -& -& -& -& 1.0& -& SCD\\
3-24& 2.50& 1.26& 20.00& 1.27& 19.27& 1.22& 13.59& 0.86& 11.01& 5.52& 0.11& 5.41& 0.96& 0.59& 0.02& AL-CM\\
3-25& 2.50& 1.26& 25.00& 1.59& 24.42& 1.55& 20.24& 1.28& 11.01& 5.16& 0.06& 5.10& 0.89& 0.78& 0.07& AL-CA\\
3-26& 2.50& 1.26& 30.00& 1.90& 29.46& 1.87& 26.17& 1.66& 11.01& 4.52& 0.03& 4.48& 0.78& 0.89& 0.19& AL-CA\\
3-27& 2.50& 1.26& 35.00& 2.22& 34.47& 2.19& 31.78& 2.02& 11.00& 3.62& 0.01& 3.61& 0.63& 0.96& 0.34& TAL-CA\\
3-28& 2.50& 1.26& 40.00& 2.54& 39.47& 2.50& 37.21& 2.36& 11.01& 2.56& 0.00& 2.56& 0.44& 1.0& 0.53& TAL-CE\\
3-29& 3.00& 1.33& 20.00& 1.23& 19.20& 1.19& 13.14& 0.81& 11.20& 6.01& 0.11& 5.90& 0.96& 0.59& 0.02& AL-CM\\
3-30& 3.00& 1.33& 25.00& 1.54& 24.38& 1.50& 19.94& 1.23& 11.21& 5.68& 0.07& 5.61& 0.91& 0.74& 0.07& AL-CA\\
3-31& 3.00& 1.33& 30.00& 1.85& 29.43& 1.82& 25.93& 1.60& 11.22& 5.01& 0.04& 4.97& 0.8& 0.85& 0.17& AL-CA\\
3-32& 3.00& 1.33& 35.00& 2.16& 34.46& 2.13& 31.58& 1.95& 11.21& 4.06& 0.02& 4.04& 0.65& 0.93& 0.33& AL-CA\\
3-33& 3.00& 1.33& 40.00& 2.47& 39.47& 2.44& 37.05& 2.29& 11.22& 2.93& 0.00& 2.93& 0.47& 1.0& 0.51& TAL-CE\\
3-34& 3.00& 1.33& 50.00& 3.09& 49.45& 3.05& 47.67& 2.94& 11.21& -& -& -& -& 1.0& -& SCD\\
\hline
\end{tabular}
\end{table*}

\begin{table*}
\caption{
A summary of collision parameters and results for head-on collisions between a $10.50\,\mathrm{M}_{\oplus}$ target ($M_\mathrm{t}^{\mathrm{core}} = 7.07 \, \mathrm{M}_{\oplus}$, $M_\mathrm{t}^{\mathrm{atmos}} = 3.43 \, \mathrm{M}_{\oplus}$) with a mantle surface radius of $1.60\,\mathrm{R}_{\oplus}$, and an atmosphere scale height of $0.70\,\mathrm{R}_{\oplus}$, and projectiles of mass $M_{\mathrm{p}}$. Similar to Tables \ref{tab:5_1.25} and \ref{tab:3_0.27}.}
\label{tab:7_3.43}
\begin{tabular}{lccccccccccccccccc}
\hline
ID&$M_{\mathrm{p}}$ &$R_{p}$ &$v_{\mathrm{imp}}^{\mathrm{pred}}$ &$\frac{v_{\mathrm{imp}}^{\mathrm{pred}}}{v_{\mathrm{esc}}}$ &$v_{\mathrm{imp}}^{\mathrm{meas}}$ &$\frac{v_{\mathrm{imp}}^{\mathrm{meas}}}{v_{\mathrm{esc}}}$ &$v_{\mathrm{init}}$ &$\frac{v_{\mathrm{init}}}{v_{\mathrm{esc}}}$ &$S$ &$M_{\mathrm{LR}}$ &$M_{\mathrm{LR}}^{\mathrm{atmos}}$ &$M_{\mathrm{LR}}^{\mathrm{core}}$ &$\frac{M_{\mathrm{LR}}}{M_{\mathrm{tot}}}$ & $X_{\mathrm{loss}}^{\mathrm{atmos}}$ & $X_{\mathrm{loss}}^{\mathrm{c\&m}}$ & Category \\
&$\mathrm{M}_{\oplus}$ &$\mathrm{R}_{\oplus}$ &$\mathrm{km}\,\mathrm{s}^{-1}$ & &$\mathrm{km}\,\mathrm{s}^{-1}$ & &$\mathrm{km}\,\mathrm{s}^{-1}$ & &$\mathrm{R}_{\oplus}$ &$\mathrm{M}_{\oplus}$ &$\mathrm{M}_{\oplus}$ &$\mathrm{M}_{\oplus}$& & & & \\

\hline
7-0& 1.01& 0.95& 30.00& 1.51& 27.76& 1.39& 21.07& 1.06& 13.54& 11.20& 3.12& 8.08& 0.97& 0.09& 0.0& AL-CM\\
7-1& 1.01& 0.95& 40.00& 2.01& 37.18& 1.87& 33.82& 1.70& 13.54& 10.81& 2.73& 8.08& 0.94& 0.2& 0.0& AL-CM\\
7-2& 1.01& 0.95& 50.00& 2.51& 46.47& 2.33& 45.21& 2.27& 13.55& 10.22& 2.15& 8.07& 0.89& 0.37& 0.0& AL-CM\\
7-3& 1.01& 0.95& 55.00& 2.76& 51.12& 2.57& 50.68& 2.54& 13.57& 9.85& 1.79& 8.07& 0.86& 0.48& 0.0& AL-CM\\
7-4& 1.01& 0.95& 60.00& 3.01& 55.99& 2.81& 56.07& 2.81& 13.57& 9.48& 1.42& 8.06& 0.82& 0.59& 0.0& AL-CM\\
7-5& 1.01& 0.95& 65.00& 3.26& 60.57& 3.04& 61.39& 3.08& 13.56& 9.05& 1.03& 8.01& 0.79& 0.7& 0.01& AL-CM\\
7-6& 1.01& 0.95& 70.00& 3.51& 65.18& 3.27& 66.66& 3.35& 13.58& 8.55& 0.70& 7.85& 0.74& 0.8& 0.03& AL-CM\\
7-7& 1.01& 0.95& 80.00& 4.01& 74.50& 3.74& 77.10& 3.87& 13.55& 6.96& 0.24& 6.72& 0.6& 0.93& 0.17& AL-CE\\
7-8& 2.02& 1.17& 30.00& 1.46& 28.16& 1.37& 21.11& 1.03& 14.17& 11.85& 2.76& 9.09& 0.95& 0.2& 0.0& AL-CM\\
7-9& 2.02& 1.17& 40.00& 1.95& 37.80& 1.84& 33.85& 1.65& 14.19& 11.03& 1.96& 9.08& 0.88& 0.43& 0.0& AL-CM\\
7-10& 2.02& 1.17& 50.00& 2.44& 47.32& 2.31& 45.23& 2.21& 14.18& 10.15& 1.16& 8.98& 0.81& 0.66& 0.01& AL-CM\\
7-11& 2.02& 1.17& 55.00& 2.68& 52.02& 2.54& 50.70& 2.47& 14.19& 9.42& 0.76& 8.66& 0.75& 0.78& 0.05& AL-CM\\
7-12& 2.02& 1.17& 60.00& 2.93& 56.64& 2.76& 56.09& 2.74& 14.19& 8.40& 0.40& 7.99& 0.67& 0.88& 0.12& AL-CA\\
7-13& 2.02& 1.17& 65.00& 3.17& 61.47& 3.00& 61.41& 3.00& 14.16& 7.14& 0.19& 6.95& 0.57& 0.94& 0.24& AL-CE\\
7-14& 2.02& 1.17& 70.00& 3.42& 66.26& 3.23& 66.67& 3.25& 14.21& 5.47& 0.02& 5.45& 0.44& 0.99& 0.4& TAL-CE\\
7-15& 2.02& 1.17& 80.00& 3.90& 75.81& 3.70& 77.11& 3.76& 14.19& 1.65& 0.00& 1.65& 0.13& 1.0& 0.82& TAL-CE\\
7-16& 3.03& 1.32& 30.00& 1.43& 28.33& 1.35& 20.89& 0.99& 14.62& 12.58& 2.49& 10.09& 0.93& 0.27& 0.0& AL-CM\\
7-17& 3.03& 1.32& 40.00& 1.90& 38.07& 1.81& 33.71& 1.60& 14.61& 11.63& 1.58& 10.05& 0.86& 0.54& 0.01& AL-CM\\
7-18& 3.03& 1.32& 50.00& 2.38& 47.68& 2.27& 45.13& 2.15& 14.64& 10.07& 0.78& 9.28& 0.74& 0.77& 0.08& AL-CA\\
7-19& 3.03& 1.32& 55.00& 2.62& 52.50& 2.50& 50.61& 2.41& 14.65& 8.87& 0.42& 8.45& 0.66& 0.88& 0.16& AL-CA\\
7-20& 3.03& 1.32& 60.00& 2.85& 57.25& 2.72& 56.00& 2.66& 14.64& 7.31& 0.17& 7.15& 0.54& 0.95& 0.29& TAL-CA\\
7-21& 3.03& 1.32& 65.00& 3.09& 62.05& 2.95& 61.33& 2.92& 14.65& 5.02& 0.01& 5.01& 0.37& 1.0& 0.5& TAL-CE\\
7-22& 3.03& 1.32& 70.00& 3.33& 66.78& 3.18& 66.61& 3.17& 14.62& 3.05& 0.00& 3.05& 0.23& 1.0& 0.7& TAL-CE\\
7-23& 3.03& 1.32& 80.00& 3.80& 76.41& 3.63& 77.05& 3.66& 14.64& -& -& -& -& 1.0& -& SCD\\
7-24& 4.04& 1.43& 30.00& 1.39& 28.37& 1.32& 20.53& 0.95& 14.94& 13.41& 2.31& 11.10& 0.92& 0.33& 0.0& AL-CM\\
7-25& 4.04& 1.43& 40.00& 1.86& 38.19& 1.77& 33.49& 1.55& 14.95& 12.39& 1.43& 10.96& 0.85& 0.58& 0.01& AL-CM\\
7-26& 4.04& 1.43& 50.00& 2.32& 47.91& 2.22& 44.96& 2.09& 14.97& 10.29& 0.63& 9.66& 0.71& 0.82& 0.13& AL-CA\\
7-27& 4.04& 1.43& 55.00& 2.55& 52.72& 2.45& 50.46& 2.34& 14.94& 8.81& 0.30& 8.51& 0.61& 0.91& 0.23& AL-CA\\
7-28& 4.04& 1.43& 60.00& 2.79& 57.57& 2.67& 55.87& 2.59& 14.96& 6.32& 0.06& 6.25& 0.43& 0.98& 0.44& TAL-CE\\
7-29& 4.04& 1.43& 65.00& 3.02& 62.36& 2.89& 61.20& 2.84& 14.98& 4.01& 0.00& 4.01& 0.28& 1.0& 0.64& TAL-CE\\
7-30& 4.04& 1.43& 70.00& 3.25& 67.15& 3.12& 66.49& 3.09& 14.98& 1.75& 0.00& 1.75& 0.12& 1.0& 0.84& TAL-CE\\
7-31& 4.04& 1.43& 80.00& 3.71& 76.75& 3.56& 76.95& 3.57& 14.99& -& -& -& -& 1.0& -& SCD\\
7-32& 5.05& 1.52& 30.00& 1.36& 28.40& 1.29& 20.11& 0.91& 15.21& 14.36& 2.25& 12.11& 0.92& 0.34& 0.0& AL-CM\\
7-33& 5.05& 1.52& 40.00& 1.82& 38.24& 1.74& 33.23& 1.51& 15.22& 13.19& 1.38& 11.81& 0.85& 0.6& 0.03& AL-CM\\
7-34& 5.05& 1.52& 50.00& 2.27& 48.01& 2.18& 44.77& 2.03& 15.21& 10.73& 0.55& 10.18& 0.69& 0.84& 0.16& AL-CA\\
7-35& 5.05& 1.52& 55.00& 2.50& 52.85& 2.40& 50.29& 2.28& 15.24& 9.01& 0.22& 8.79& 0.58& 0.94& 0.27& AL-CA\\
7-36& 5.05& 1.52& 60.00& 2.72& 57.69& 2.62& 55.71& 2.53& 15.25& 6.17& 0.02& 6.16& 0.4& 0.99& 0.49& TAL-CE\\
7-37& 5.05& 1.52& 65.00& 2.95& 62.53& 2.84& 61.07& 2.77& 15.25& 3.78& 0.00& 3.77& 0.24& 1.0& 0.69& TAL-CE\\
7-38& 5.05& 1.52& 70.00& 3.18& 67.35& 3.06& 66.37& 3.01& 15.21& 1.65& 0.00& 1.65& 0.11& 1.0& 0.86& TAL-CE\\
7-39& 5.05& 1.52& 80.00& 3.63& 77.01& 3.50& 76.84& 3.49& 15.25& -& -& -& -& 1.0& -& SCD\\
7-40& 6.06& 1.60& 30.00& 1.33& 28.39& 1.26& 19.67& 0.87& 15.46& 15.33& 2.22& 13.11& 0.93& 0.35& 0.0& AL-CM\\
7-41& 6.06& 1.60& 40.00& 1.78& 38.27& 1.70& 32.97& 1.47& 15.46& 14.14& 1.40& 12.74& 0.85& 0.59& 0.03& AL-CM\\
7-42& 6.06& 1.60& 50.00& 2.22& 48.04& 2.14& 44.57& 1.98& 15.47& 11.40& 0.51& 10.89& 0.69& 0.85& 0.17& AL-CA\\
7-43& 6.06& 1.60& 55.00& 2.44& 52.96& 2.35& 50.12& 2.23& 15.46& 9.47& 0.20& 9.27& 0.57& 0.94& 0.29& AL-CA\\
7-44& 6.06& 1.60& 60.00& 2.67& 57.82& 2.57& 55.56& 2.47& 15.48& 6.35& 0.01& 6.34& 0.38& 1.0& 0.52& TAL-CE\\
7-45& 6.06& 1.60& 65.00& 2.89& 62.62& 2.78& 60.92& 2.71& 15.46& 3.90& 0.00& 3.90& 0.24& 1.0& 0.7& TAL-CE\\
7-46& 6.06& 1.60& 70.00& 3.11& 67.50& 3.00& 66.23& 2.94& 15.49& 1.78& 0.00& 1.78& 0.11& 1.0& 0.86& TAL-CE\\
7-47& 6.06& 1.60& 80.00& 3.56& 77.14& 3.43& 76.72& 3.41& 15.50& -& -& -& -& 1.0& -& SCD\\
7-48& 7.07& 1.67& 30.00& 1.31& 28.38& 1.24& 19.18& 0.84& 15.67& 16.30& 2.18& 14.12& 0.93& 0.36& 0.0& AL-CM\\
7-49& 7.07& 1.67& 40.00& 1.74& 38.30& 1.67& 32.68& 1.42& 15.67& 15.10& 1.36& 13.75& 0.86& 0.6& 0.03& AL-CM\\
7-50& 7.07& 1.67& 50.00& 2.18& 48.10& 2.10& 44.36& 1.93& 15.67& 12.19& 0.50& 11.69& 0.69& 0.85& 0.17& AL-CA\\
7-51& 7.07& 1.67& 55.00& 2.40& 52.98& 2.31& 49.93& 2.18& 15.65& 10.30& 0.19& 10.11& 0.59& 0.94& 0.29& AL-CA\\
7-52& 7.07& 1.67& 60.00& 2.61& 57.85& 2.52& 55.39& 2.41& 15.66& 7.00& 0.01& 6.99& 0.4& 1.0& 0.51& TAL-CE\\
7-53& 7.07& 1.67& 65.00& 2.83& 62.66& 2.73& 60.77& 2.65& 15.70& 4.27& 0.00& 4.26& 0.24& 1.0& 0.7& TAL-CE\\
7-54& 7.07& 1.67& 70.00& 3.05& 67.55& 2.94& 66.09& 2.88& 15.69& 2.23& 0.00& 2.23& 0.13& 1.0& 0.84& TAL-CE\\
7-55& 7.07& 1.67& 80.00& 3.49& 77.25& 3.37& 76.60& 3.34& 15.67& -& -& -& -& 1.0& -& SCD\\

\hline
\end{tabular}
\end{table*}


\bsp	
\label{lastpage}
\end{document}